\documentclass[longauth]{aa}
\usepackage{natbib}
\usepackage[varg]{txfonts}
\usepackage{graphicx}

\usepackage{booktabs}
\usepackage{multirow}
\usepackage{tabularx}
\usepackage{lscape}
\usepackage{longtable}

\newcolumntype{L}[1]{>{\raggedright\arraybackslash}p{#1}}
\newcolumntype{C}[1]{>{\centering\arraybackslash}p{#1}}
\newcolumntype{R}[1]{>{\raggedleft\arraybackslash}p{#1}}

\usepackage{color} 

\def\Msun{\ifmmode{\mathrm M_\odot}\else{$M_\odot$}\fi}

\def\mod#1{{#1}}

\begin{document}

\defcitealias{2012MNRAS.421.3127N}{N12}
\defcitealias{2013ARA&A..51..207B}{B13}
\defcitealias{2015A&A...582A..86H}{HE15}

\title{Stellar structures, molecular gas, and star formation across the PHANGS sample of nearby galaxies}

\author{M.~Querejeta\inst{1}, E.~Schinnerer\inst{2}, S.~Meidt\inst{3},  J.~Sun\inst{4}, A.~K.~Leroy\inst{4},  E.~Emsellem\inst{5,6},  R.~S.~Klessen\inst{7,8},
J.~C.~Muñoz-Mateos\inst{2}, H.~Salo\inst{9}, E.~Laurikainen\inst{9}, 
I.~Be\v{s}li\'c\inst{10}, G.~A.~Blanc\inst{11,12}, M.~Chevance\inst{13},  D.~A.~Dale\inst{14}, C.~Eibensteiner\inst{10}, C.~Faesi\inst{15}, A.~Garc\'{i}a-Rodr\'{i}guez\inst{1}, S.~C.~O.~Glover\inst{7}, K.~Grasha\inst{16}, J.~Henshaw\inst{2}, C.~Herrera\inst{17}, A.~Hughes\inst{18,19}, K.~Kreckel\inst{13}, J.~M.~D.~Kruijssen\inst{13}, D.~Liu\inst{2}, E.~J.~Murphy\inst{20}, H.-A.~Pan\inst{2}, J.~Pety\inst{17,21}, A.~Razza\inst{12}, E.~Rosolowsky\inst{22}, T.~Saito\inst{2}, A.~Schruba\inst{23}, A.~Usero\inst{1}, E.~J.~Watkins\inst{18}, T.~G.~Williams\inst{2}
}

\institute{Observatorio Astron{\'o}mico Nacional (IGN), C/ Alfonso XII 3, E-28014 Madrid, Spain, \email{m.querejeta@oan.es}
\and  Max-Planck-Institut f\"{u}r Astronomie, K\"{o}nigstuhl 17, D-69117 Heidelberg, Germany 
\and Sterrenkundig Observatorium, Universiteit Gent, Krijgslaan 281 S9, B-9000 Gent, Belgium 
\and Department of Astronomy, The Ohio State University, 140 West 18th Ave, Columbus, OH 43210, USA 
\and European Southern Observatory, Karl-Schwarzschild-Stra{\ss}e 2, D-85748 Garching, Germany  
\and Univ Lyon, Univ Lyon1, ENS de Lyon, CNRS, Centre de Recherche Astrophysique de Lyon UMR5574, F-69230 Saint-Genis-Laval, France 
\and Universit\"{a}t Heidelberg, Zentrum f\"{u}r Astronomie, Albert-Ueberle-Stra{\ss}e 2, D-69120 Heidelberg, Germany  
\and Universit\"{a}t Heidelberg, Interdisziplin\"{a}res Zentrum f\"{u}r Wissenschaftliches Rechnen, INF 205, D-69120 Heidelberg, Germany 
\and Space Physics and Astronomy Research Unit, University of Oulu, Pentti Kaiteran katu 1, FI-90014, Finland 
\and Argelander-Institut f\"ur Astronomie, Universit\"{a}t Bonn, Auf dem  H\"ugel 71, D-53121 Bonn, Germany 
\and The Observatories of the Carnegie Institution for Science, 813 Santa Barbara Street, Pasadena, CA 91101, USA 
\and Departamento de Astronomía, Universidad de Chile, Casilla 36-D, Santiago, Chile
\and Zentrum f\"{u}r Astronomie der Universit\"{a}t Heidelberg, Astronomisches Rechen-Institut, M\"{o}nchhofstra{\ss}e 12-14, D-69120 Heidelberg, Germany   
\and Department of Physics \& Astronomy, University of Wyoming, Laramie, WY 82071 
\and Department of Astronomy, University of Massachusetts Amherst, 710 North Pleasant St., Amherst, MA 01003  
\and Research School of Astronomy and Astrophysics, Australian National University, Canberra, ACT 2611, Australia 
\and IRAM, 300 rue de la Piscine, F-38406 Saint Martin d’H\`eres, France
\and CNRS, IRAP, 9 Av. du Colonel Roche, BP 44346, F-31028 Toulouse cedex 4, France 
\and Universit\'{e} de Toulouse, UPS-OMP, IRAP, F-31028 Toulouse cedex 4, France
\and NRAO, 520 Edgemont Road, Charlottesville, VA 22903
\and Sorbonne Universit\'e, Observatoire de Paris, Universit\'e PSL, CNRS, LERMA, F-75005 Paris, France
\and 4-183 CCIS, University of Alberta, Edmonton, Alberta, Canada
\and Max-Planck-Institut f\"{u}r extraterrestrische Physik, Giessenbachstra{\ss}e 1, D-85748 Garching, Germany
}

\date{Received ..... / Accepted .....}

\abstract {We identify stellar structures in the PHANGS sample of 74 nearby galaxies and construct \mod{morphological masks of sub-galactic environments} based on {\it Spitzer} $3.6$\,$\mu$m images. At the simplest level, we distinguish centres, bars, spiral arms, interarm regions, and discs without strong spirals. Slightly more sophisticated masks include rings and lenses, which are \mod{publicly released but} not explicitly used in this paper. We examine trends with environment in the molecular gas content, star formation rate, and depletion time \mod{using PHANGS--ALMA \mbox{CO(2--1)} intensity maps and tracers of star formation.} The interarm regions and discs without strong spirals clearly dominate in area, whereas molecular gas and star formation are quite evenly distributed among the five basic environments. We reproduce the molecular Kennicutt--Schmidt relation with a slope \mod{compatible with unity within the uncertainties and without a significant slope differences among environments.} \mod{In contrast to what has been suggested by early studies,} we find that bars are not always deserts devoid of gas and star formation, but instead they show large diversity. Similarly, spiral arms do not account for most of the gas and star formation in disc galaxies, and they do not have shorter depletion times than the interarm regions. Spiral arms accumulate gas and star formation, without systematically boosting the star formation efficiency. Centres harbour remarkably high surface densities and on average shorter depletion times than other environments. \mod{Centres of barred galaxies show higher surface densities and wider distributions compared to the outer disc; yet, depletion times are similar to unbarred galaxies, suggesting highly intermittent periods of star formation when bars episodically drive gas inflow, without enhancing the central star formation efficiency permanently.} In conclusion, we provide quantitative evidence that stellar structures in galaxies strongly affect the organisation of molecular gas and star formation, but their impact on star formation efficiency is more subtle.}


\titlerunning{Morphological environments in PHANGS}
\authorrunning{M.~Querejeta et al.}

\maketitle 
\section{Introduction} 
\label{Sec:introduction}

Galaxies in the local Universe display a wealth of morphological structures including bars, rings, and spiral arms. These features are the result of the evolution driven both by internal and external mechanisms, and they hold key information to unravel the assembly history of galaxies across cosmic time. Specifically, some of these morphological structures have been argued to play a pivotal role in the so-called secular evolution of galaxies \citep[see e.g.][for a review]{2004ARA&A..42..603K}.
For instance, bars tend to drive gas inwards and can potentially feed an active nucleus \citep[e.g.][]{1987MNRAS.225..653S,1992MNRAS.259..345A,1999MNRAS.304..475M,2003ASPC..290..411C,2006LNP...693..143J}. The accumulation of gas in spiral arms or rings can impact the properties of molecular gas and possibly its ability to form new stars \citep[e.g.][]{2011MNRAS.417.1318D,2012A&A...542A..39G,2013ApJ...769..100S,2013ApJ...779...42S,2017A&A...603A.113S,2020MNRAS.497.5024S}.
Another example is the emergence of a spheroidal stellar component in a disc galaxy, which has been argued to quench star formation \citep[the so-called morphological quenching or dynamical suppression; ][]{2009ApJ...707..250M,2020MNRAS.495..199G,2021MNRAS.500.2000G}.
All in all, these stellar structures shape gas and star formation in galaxies and play a major role in galaxy evolution.

Far from being rare, bars, rings, and spiral arms are ubiquitous in the present-day Universe. Nearly two out of three galaxies in the local Universe host stellar bars \citep[e.g.][]{1963ApJS....8...31D,2000AJ....119..536E,2007ApJ...657..790M,2011MNRAS.411.2026M} with some dependence on the morphological type \citep{2015ApJS..217...32B,2016A&A...587A.160D}. On average, around 35\% of nearby galaxies host inner rings; as much as ${\sim}50$\% for early-type spirals (Sa-Sbc), with a sharp decline down to ${\sim}15$\% for late-type spirals \citep[Sc-Sm;][]{2014A&A...562A.121C,2015ApJS..217...32B}. Outer rings appear somewhat less common \citep[e.g.\ 16\% measured by ][]{2014A&A...562A.121C}, but their detection might be compromised by the lower surface brightness of galaxy outskirts. In the same vein, roughly two thirds of nearby galaxies display some kind of spiral structure \citep[e.g.][]{2010ApJS..186..427N,2013MNRAS.435.2835W,2015ApJS..217...32B}, ranging from grand-design spirals, with two long symmetric arms, to multi-armed and flocculent spirals, where the spiral segments become increasingly weaker, shorter, and less distinct.

There is mounting evidence that the properties of molecular gas and star formation are regulated by local galactic environment. Indeed, galactic structures have an influence on molecular gas probability distribution functions \citep[PDFs; e.g.][]{2013ApJ...779...44H,2018ApJ...854...90E,2021ApJ...913..113M}. Galactic environment can also affect the properties and evolution of giant molecular clouds \citep[GMCs; e.g.][]{2014ApJ...784....3C,2016IAUS..315...30H,2019ApJ...883....2S,2020MNRAS.493.5045M,2020NatAs...4.1064H,2020MNRAS.493.2872C,2020ApJ...901L...8S,2021MNRAS.502.1218R}. In particular, stellar structures and local dynamical environment have been recognised as factors that modulate star formation, either enhancing or suppressing it \citep[e.g.][]{2013ApJ...779...45M,2015MNRAS.454.3299R,2016ApJ...818...69M,2018ApJ...853..149S}.
In this paper, we study molecular gas and star formation across a representative sample of nearby galaxies; specifically, we try to understand how stellar environments orchestrate the distribution of molecular gas and its ability to form stars.

The near infrared (NIR) has been exploited as a privileged wavelength range to identify stellar structures, given that it is minimally affected by dust extinction and shows only weak variations in the stellar mass-to-light ratio \mod{ \citep[e.g.][]{2007ApJ...657..790M,2010PASP..122.1397S,2011MNRAS.418.1452L,2016MNRAS.455.3911D}. }
The stellar structures visible in the NIR constitute the fossil record of the processes that shaped galaxies to their current state. They are also a proxy for different dynamical regions, each associated with a complex backbone made of stellar and gaseous orbits within a time-varying potential. Each type of structure (e.g.\ bars, rings, spiral arms) can be interpreted via specific tracers \citep[e.g.][]{1992MNRAS.259..328A}, specific torque maps or gas flows \citep[e.g.][]{2005A&A...441.1011G,2009ApJ...692.1623H,2016A&A...588A..33Q,2016ApJ...823...85L}, and sometimes connected with density waves such as spiral arms \citep{1964ApJ...140..646L,1989ApJ...343..602E,1996ssgd.book.....B} and their associated resonances such as rings \citep[e.g.][]{1996FCPh...17...95B,2014A&A...562A.121C,2017MNRAS.471.4027B}.

Recent observations with deep NIR exposures and large samples have revealed increasingly diverse stellar structures in local galaxies \citep[e.g.][]{2010ApJS..190..147B,2015ApJS..217...32B,2019MNRAS.486.1995S}. The {\it Spitzer} Survey of Stellar Structures in Galaxies \citep[S$^4$G; ][]{2010PASP..122.1397S} has made a significant contribution to this field by surveying a large set of galaxies in the NIR. With $2352$ galaxies, S$^4$G constitutes the largest detailed inventory of stellar structures observed so far in the nearby Universe. Here we use S$^4$G \mod{and other NIR observations to construct} a detailed, homogeneous set of \mod{morphological masks of sub-galactic environments} for galaxies in PHANGS.

The PHANGS\footnote{Physics at High Angular resolution in Nearby GalaxieS; \url{http://www.phangs.org}} project involves a set of surveys of nearby galaxies conducted at ${\lesssim} 1''$ resolution to understand the details of the star formation process in galaxies. One of the main goals of PHANGS is to underpin the environmental dependence of the star formation cycle, observing molecular gas at cloud scales, its collapse to form stars, and the different forms of feedback associated with star formation. To that aim, three large programmes on ALMA \citep{2021arXiv210407739L}, MUSE \citep{emsellem21}, and HST \citep{2021arXiv210102855L}, as well as a number of smaller focused programmes, provide complementary viewpoints on the evolutionary stages involved in the process of star formation. The large census of GMCs, {\sc H\,ii} regions, and stellar clusters revealed by this coordinated effort should provide robust statistics on the star formation cycle for different environments in a diverse sample of nearby star-forming galaxies.

To make best use of these data, PHANGS requires a homogeneous definition of galactic environments for all targets. This will allow rigorous comparative analysis and enable measurements that assess the impact of environment on star formation and feedback processes. In this paper, we present a first approach to identifying galactic environments based on NIR photometry that traces stellar structures. We construct a set of 2D environmental masks relying on {\it Spitzer} IRAC $3.6$\,$\mu$m images, which are now homogeneously available for the whole PHANGS sample at a resolution of ${\sim} 1.7\arcsec$ \mod{(either from S$^4$G or from other archival or new \textit{Spitzer} observations)}. This is usually sufficient to resolve the stellar structures in which we are primarily interested (e.g.\ stellar bars and spiral arms) and is not far from the typical PHANGS--ALMA resolution of ${\sim} 1\arcsec$. 

PHANGS--ALMA has mapped a set of galaxy discs that host molecular gas and star formation, but has also shown that not all of this molecular gas is instantaneously associated with massive stars
\citep{2019ApJ...887...49S,pan21}. At the highest resolution of the PHANGS--ALMA survey (${\sim}$100\,pc), corresponding to the sizes of massive GMCs, a large fraction of the sight lines are associated only with CO emission but not with H$\alpha$ emission, which traces star formation. 
This motivates the main scientific question of this paper: How is molecular gas and star formation organised across the PHANGS sample of nearby galaxies? What role does galactic environment play?

This paper is structured as follows. In Sect.~\ref{Sec:data} we describe \mod{the galaxy sample (Sect.~\ref{Sec:sample}),} the observations from  {\it Spitzer} (Sect.~\ref{Sec:Spitzer}), ALMA (Sect.~\ref{Sec:PHANGS--ALMA}), and star formation tracers (Sect.\,\ref{Sec:SFRs}). Sect.~\ref{Sec:masks} introduces our definition of environments and how we construct the environmental masks for PHANGS. Sect.~\ref{Sec:results} presents the main results of the paper. We discuss our results in the context of previous observations in Sect.~\ref{Sec:discussion}, and we close the paper with a summary in Sect.~\ref{Sec:concl}.

\begin{figure*}[t]
\begin{center}
\includegraphics[trim=0 0 0 0, clip,width=1.0\textwidth]{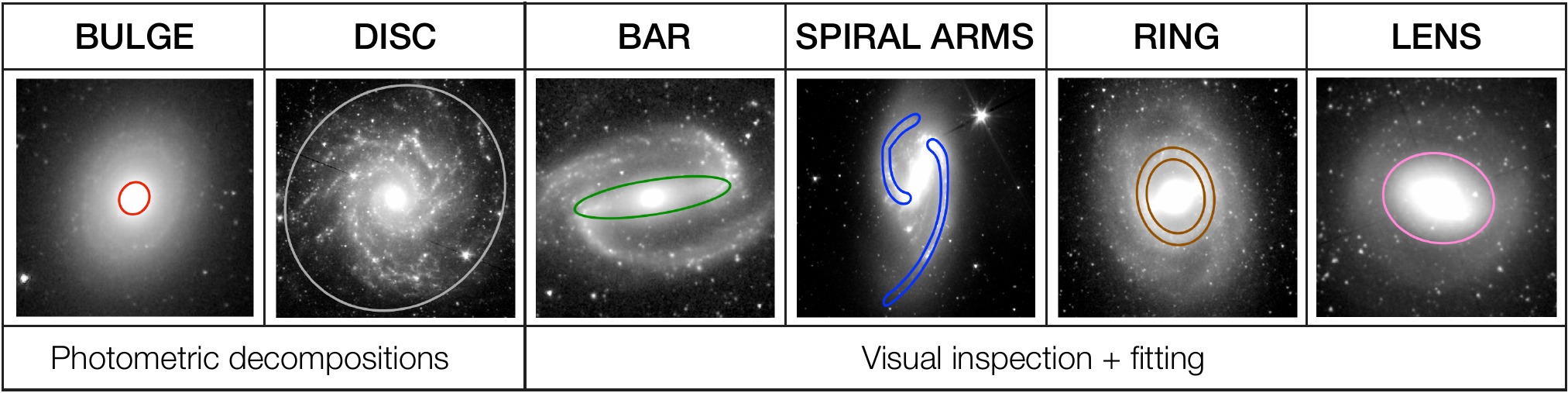}
\end{center}
\caption{\textit{Spitzer} IRAC $3.6$\,$\mu$m images illustrating the morphological structures considered in our environmental masks. Sect.~\ref{Sec:masks} explains the mask construction scheme in detail.
Bulges and discs are defined on photometric decompositions of near-infrared images (\citealt{2015ApJS..219....4S} for S$^4$G galaxies; \citealt{2004MNRAS.355.1251L} or new fits otherwise). The sizes of bars, rings and lenses are defined visually and their ellipticity is measured via ellipse fitting (\citealt{2015A&A...582A..86H} for S$^4$G, with additional measurements from the literature). Spiral arms are identified as peaks on unsharp-masked $3.6$\,$\mu$m images followed by log-spiral fits in polar coordinates, the width being assigned based on CO emission (\citealt{2015A&A...582A..86H} and new measurements).
The galaxies shown are, from left to right, NGC\,2775, NGC\,628, NGC\,1300, NGC\,3627, NGC\,3351 and NGC\,4457; they are all displayed using an arcsinh stretch.}
\label{fig:intro}
\end{figure*}

\section{Data} 
\label{Sec:data}

In this paper, we study the main PHANGS--ALMA sample of 74 nearby galaxies, made up mostly of star-forming spiral discs, \mod{as explained in Sect.~\ref{Sec:sample}}. For all those galaxies, we have gathered NIR {\it Spitzer} IRAC data, as described in Sect.\,\ref{Sec:Spitzer}. Next, Sect.\,\ref{Sec:PHANGS--ALMA} describes the ALMA observations, tracing molecular gas, while our strategy to measure star formation is presented in Sect.\,\ref{Sec:SFRs}.

\subsection{\mod{Sample}} 
\label{Sec:sample}

\mod{In this paper, we focus on the nominal PHANGS--ALMA sample of 74 galaxies, which covers most of the nearby, massive, star-forming galaxies with moderate inclinations selected to have distances out to $D \approx 17$\,Mpc and to be visible to ALMA. The list of galaxies can be found in Table~\ref{table:sample} along with some basic properties; a comprehensive description of the sample can be found in \citet{2021arXiv210407739L}.}

\mod{PHANGS--ALMA galaxies closely follow the $z=0$ `main  sequence' of star-forming galaxies, with specific star formation rates above ${\rm SFR}/M_\star > 10^{-11}$\,yr$^{-1}$. The overwhelming majority of the sample corresponds to spiral galaxies, including both early- and late-type spirals, and also contains a handful of lenticular (S0) and irregular galaxies. The sample spans stellar masses in the range $9.25 \lesssim \log(M_\star / M_\odot) \lesssim 11.25$. The PHANGS--ALMA sample grows up to $90$ galaxies if we include a number of extensions from other programmes, but those are not considered in this paper \citep[see][for details]{2021arXiv210407739L}. PHANGS--ALMA includes galaxies that were excluded from S$^4$G (nearly always due to the S$^4$G cut in Galactic latitude $|b| > 30^\circ$), but we have assembled archival or new IRAC observations for all of them.}

\subsection{{\it Spitzer} IRAC observations} 
\label{Sec:Spitzer}

\subsubsection{Products from the S$^4$G survey} 

We identify stellar structures using $3.6$\,$\mu$m images from the {\it Spitzer} Space Telescope, the shortest wavelength channel of the IRAC camera \citep{2004ApJS..154...10F}. The majority of the galaxies in our sample ($60$ out of $74$ galaxies) were mapped by the S$^4$G survey (\citealt{2010PASP..122.1397S}); for those, we rely on the products publicly released on the NASA/IPAC Infrared Science Archive\footnote{\url{https://irsa.ipac.caltech.edu/data/SPITZER/S4G}} (IRSA).

The final {\it Spitzer} IRAC $3.6$\,$\mu$m images for these targets achieve a depth of $\mu_{3.6\,\mu \rm m}\,{\rm (AB)} \sim 27$\,mag\,arcsec$^{-2}$ (equivalent to a stellar surface density of ${\sim}1$\,$M_\odot$\,pc$^{-2}$). The {\it Spitzer} IRAC point spread function (PSF) has a complex structure with an approximate FWHM size of $1.7\arcsec$ \citep{2010PASP..122.1397S}. The images were processed using the S$^4$G pipeline, and here we rely on the location of galaxy centres determined within Pipeline~3 \citep{2015ApJS..219....3M} and the GALFIT \citep{2002AJ....124..266P,2010AJ....139.2097P} photometric decompositions from Pipeline~4 \citep{2015ApJS..219....4S}.

\subsubsection{Additional archival and new observations} 

Ten galaxies in the main PHANGS--ALMA sample were not covered by S$^4$G but had ancillary $3.6$\,$\mu$m observations available from IRSA, either from the SINGS survey \citep{2003PASP..115..928K} or from other individual observations. For those, we downloaded the SINGS products from IRSA, when available, or the {\it Spitzer} Enhanced Imaging Products otherwise. The latter are Super Mosaics where contiguous Astronomical Observation Requests (AORs) were grouped and reduced together using an automated pipeline on the {\it Spitzer} MOsaicker and Point source EXtractor (MOPEX) package\footnote{\url{https://irsa.ipac.caltech.edu/data/SPITZER/docs/dataanalysistools/tools/mopex}}. We refer the reader to the online documentation on the {\it Spitzer} Enhanced Imaging Products for more details\footnote{\url{https://irsa.ipac.caltech.edu/data/SPITZER/Enhanced/SEIP/docs/seip_explanatory_supplement_v3.pdf}}.

Four PHANGS galaxies did not have any archival {\it Spitzer} IRAC observations at $3.6$\,$\mu$m, so we obtained their imaging in Cycle~14 via a dedicated programme (pid 14033, PI J.~C.~Mu\~noz-Mateos). The galaxies NGC\,2283, NGC\,2835, NGC\,3059 and NGC\,3137 were observed between 2018-09-18 and 2019-04-03. We followed the same observing strategy as used by S$^4$G: all galaxies fit within the IRAC field of view, so we covered them with a single pointing using a small dither cycling pattern with 4~positions ($30$~second exposures at each dither position). Each target was observed with two AORs spaced by several weeks, yielding a total exposure time per pixel of 4~minutes (the same as for the SINGS IRAC imaging; \citealt{2003PASP..115..928K}), resulting in a similar depth as S$^4$G.
We reduced the data using MOPEX, starting from the corrected basic calibrated data files (CBCD files), which is the result of running the IRAC artefact correction pipeline on the native BCD files for each AOR. We then ran the MOPEX Mosaicker, initiating a `New Overlap Pipeline' and adding the CBCD files corresponding to the two epochs with their corresponding uncertainty files. We chose `Fiducial Image Frame' so that the mosaics follow the usual orientation of north up, east left. With this strategy, the image quality and characteristics are very similar to the S$^4$G maps.

Finally, we ran all the IRAC images through \url{astrometry.net} to correct for possible small astrometric offsets \citep{2010AJ....139.1782L}. We used index files $4202$ and $4203$, which are based on the Two-Micron All-Sky Survey (2MASS; \citealt{2006AJ....131.1163S}) catalogue.

\subsection{PHANGS--ALMA data} 
\label{Sec:PHANGS--ALMA}

PHANGS has uniformly mapped $^{12}$\mbox{CO(2--1)} emission across a sample of nearby star-forming galaxies with ALMA, and the survey is presented in detail in \citet{2021arXiv210407739L}. Here we focus on the main PHANGS--ALMA sample of 74 nearby galaxies, excluding extensions from other programmes. Our typical ALMA angular resolution of ${\sim}1\arcsec$ translates into a physical scale of $50{-}150$\,pc. Our observations combine ALMA 12\,m-array, 7\,m-array and total-power \mod{(TP)} single-dish data in order to recover emission from all spatial scales. \mod{The ALMA field of view, to which we restricted our measurements, was designed to cover the star-forming part of the galactic disc (where \textit{WISE} $12$\,$\mu$m surface brightness exceeds $0.5$\,MJy~sr$^{-1}$ at $7.5\arcsec$ resolution). This corresponds to a coverage typically extending out to ${\sim}R_{25}$ (the median $R_{\rm max}/R_{25}$ is $0.99$), with full azimuthal coverage usually out to ${\sim}R_{25}/2$ (the median $R_{\rm max}^{\rm uniform}/R_{25}$ is $0.55$). This area typically encompasses ${\sim} 70{-}90\%$ of the total CO emission \citep{2021ApJS..255...19L}.}

\mod{We measured CO integrated intensity as the zeroth-order moment maps at the native ALMA resolution. These maps were computed using very simple and inclusive masks, constructed to span the full velocity coverage of the galaxy along each line of sight (the specific velocity integration window for each galaxy is listed in Table~\ref{table:sample}). This ensures more uniform noise and high completeness at the expense of signal to noise. However, since we are working at relatively low resolution, where signal to noise is not a major limiting factor for PHANGS--ALMA, we prefer these simple and inclusive masks (which are even less restrictive than the `broad' masks presented in \citealt{2021ApJS..255...19L}).  We use the PHANGS data cubes publicly released in mid-2021 (PHANGS--ALMA version~4.0).}

We convert from the observed CO integrated intensities (in K\,km\,s$^{-1}$) to surface densities (in $M_\odot$\,pc$^{-2}$) as $\Sigma_{\rm mol} = \alpha_{\rm CO}\, R_{21}^{-1}\, I_{\rm CO}^{21} \cos(i)$, where $i$ is the inclination of the disc. We adopt a line ratio of $R_{21}=0.65$ \citep{2013AJ....146...19L,denBrok21}. Our preferred approach is to use a metallicity-dependent $\alpha_{\rm CO}$ conversion factor as detailed in \citet{2020ApJ...892..148S}:

\begin{equation}
    \begin{split}
    \alpha_{\rm CO}^{\rm PHANGS} = 4.35\,{\rm M_\odot\,pc^{-2}\,(K\,km\,s^{-1})^{-1}} \times Z'^{-1.6}~,
    \end{split}
\end{equation}

\noindent
where $Z'$ is the local metallicity (normalised by the solar metallicity), a scaling introduced by \citet{2017MNRAS.470.4750A}. This local $Z'$ is estimated from the global galaxy metallicity (via the mass--metallicity relation) and a fixed radial metallicity gradient in each galaxy ($-0.1\,{\rm dex}\,R_{\rm e}^{-1}$; \citealt{2014A&A...563A..49S,2019MNRAS.484.3042S}), as explained in \citet{2020ApJ...892..148S}. This means that our adopted $\alpha_{\rm CO}$ varies only radially, in a smooth way, and is by construction available for all positions in the galaxy, because it does not depend on the availability of local metallicity measurements.

To test how sensitive our results are to this choice of $\alpha_{\rm CO}$, we also consider alternative prescriptions for the conversion factor \mod{in Appendix~\ref{sec:Appendix1} (and Table~\ref{table:KSfits}).}
We investigate the possibility of a constant $\alpha_{\rm CO}^{\rm MW} = 4.35$\,$M_\odot$\,pc$^{-2}$\,(K\,km\,s$^{-1})^{-1}$, which is the Galactic value recommended by \citet{2013ARA&A..51..207B}. Following \citet{2020ApJ...892..148S}, we also take the prescription of \citet[][hereafter \citetalias{2012MNRAS.421.3127N}]{2012MNRAS.421.3127N}, based on their Eq.\,11, which includes a dependence on metallicity and flux-weighted CO intensity. We also consider the conversion factor from \citet[][hereafter \citetalias{2013ARA&A..51..207B}]{2013ARA&A..51..207B}, based on their Eq.\,31, which depends on the local cloud-scale molecular gas surface density, metallicity, and the average kpc-scale disc surface density (including gas and stars). For consistency, in all of these cases we include a factor $1.36$ to account for heavy elements (\mod{\citealt{2013ARA&A..51..207B};} even though the original prescription of \citetalias{2012MNRAS.421.3127N} did not include it).

\vspace{1cm}

We adopt the distances, centre, and orientation parameters from the PHANGS sample table (release v1.6, late-2020; \citealt{2021arXiv210407739L}). The compilation of distances is described in \citet{2021MNRAS.501.3621A}, while the centre, inclination and position angle of the disc come from the CO kinematic analysis presented in \citet{2020ApJ...897..122L}. The metallicity-dependent $\alpha_{\rm CO}$ prescription also relies on the stellar mass and effective radius from the PHANGS sample table v1.6 \citep{2021arXiv210407739L}.

For most of this paper, we rely on measurements at matched kpc resolution from the PHANGS multi-wavelength database presented in \citet[][]{2020ApJ...901L...8S} and J.~Sun et al.\ (in prep.).
The PHANGS--ALMA CO field of view was divided into a regular tiling of hexagonal apertures, with a separation of $1$\,kpc in the plane of the sky between centres of adjacent hexagons. 
\mod{The measurements were performed at the centre of each hexagonal aperture by directly sampling a set of intensity maps that were convolved to a common physical resolution of $1.5$~kpc.}

\subsection{Star formation rate measurements} 
\label{Sec:SFRs}

In this paper, we work with star formation rates (SFRs) measured at two different resolutions.
A high-resolution SFR estimate is employed in Sect.\,\ref{Sec:piechart} to provide a census of star formation and molecular gas at ${\sim}1\arcsec$ resolution; this affords the advantage of completeness, as all pixels are considered and individually assigned to an environment. However, for most of the paper (Sects.\,\ref{Sec:surfdens}--\ref{Sec:contrasts}), we rely on kpc-scale measurements which should be more robust estimates of the star formation activity (with the drawback of sacrificing completeness, as we only retain kpc-scale apertures that are reliably associated with a given environment, as explained in Sect.\,\ref{Sec:surfdens}).

\mod{The highest resolution available for our SFR estimate (${\sim} 1\arcsec$) comes from a PHANGS survey of ground-based narrow-band H$\alpha$ images (A.~Razza et al.\ in prep.). A total of 65 galaxies from the PHANGS--ALMA parent sample were observed between 2016 and 2019 with the Wide Field Imager (WFI) at the MPG \mbox{2.2-metre} telescope in La Silla or with the DirectCCD camera at the du~Pont \mbox{2.5-metre} telescope in Las Campanas. As part of the survey, broad-band images were observed alongside the narrow-band ones to produce H$\alpha$ continuum-subtracted images with seeing-limited resolutions ranging from $0.6\arcsec$ to $1.3\arcsec$.}
H$\alpha$ images were corrected for filter transmission and [{\sc N\,ii}] contamination (assuming $[\textsc{N\,ii}]/\textrm{H}\alpha = 0.3$; e.g. \citealt{2019ApJ...887...49S}). In order to scale them to SFR units, the H$\alpha$ maps were convolved to $15\arcsec$ resolution and linearly combined with \textit{WISE} band~4 ($22$\,$\mu$m) from \citet{2019ApJS..244...24L}, which corrects for obscured star formation, following \citet{2007ApJ...666..870C} and assuming that $\nu L_\nu (22\,\mu {\rm m}) = \nu L_\nu (24\,\mu {\rm m})$ \citep{2013AJ....145....6J,2017ApJ...850...68C}. The resulting $15\arcsec$ resolution maps were then divided by the $15\arcsec$ H$\alpha$ map, which yields a scaling factor map that is then applied to the high-resolution (${\sim}1\arcsec$) H$\alpha$ narrow-band maps. Put another way, this method combines \textit{WISE} 22\,$\mu$m emission and H$\alpha$ to estimate the H$\alpha$ extinction at $15\arcsec$ resolution; then, it assumes that this extinction remains fixed or smooth at higher resolution. 

On the other hand, for most of the paper we rely on the SFRs measured at kpc scales on a hexagonal grid using a hybrid combination of UV and IR images as described in \citet{2020ApJ...901L...8S} and following \citet{2019ApJS..244...24L}. Specifically, we adopt a linear combination of \textit{GALEX} FUV and \textit{WISE} $22$\,$\mu$m as long as both bands are available (using NUV when FUV is not available, and relying only on \textit{WISE} $22$\,$\mu$m when we lack \textit{GALEX} observations). The coefficients that scale the luminosity in each band ($\nu L_\nu$) to SFRs follow the prescription presented in \citet{2019ApJS..244...24L}. This strategy aims to anchor our SFR estimates to the large set of nearby galaxies studied in \citet{2019ApJS..244...24L} and the SDSS catalogue of \citet{2016ApJS..227....2S,2018ApJ...859...11S}.

\section{Construction of environmental masks} 
\label{Sec:masks}

Our goal is to construct multi-layer binary masks (where each layer reflects a different structure) to capture the wealth of morphological environments present across PHANGS galaxies. In order to outline stellar structures, we rely on {\it Spitzer} $3.6$\,$\mu$m imaging (Sect.\,\ref{Sec:Spitzer}). We emphasise that these masks are purely morphological, and do not explicitly incorporate the kinematic information available from PHANGS, which could lead to other definitions of environments (as an example, the bar region could extend to the corotation of the bar).
\mod{Here we introduce some detailed environmental masks that are publicly released\footnote{\url{http://dx.doi.org/10.11570/21.0024} } and have been used in other publications \citep[e.g.][]{2021MNRAS.506..963B,2021A&A...650A.134P}. For the analysis in this paper, we focus on five basic environments as described in Sect.\,\ref{Sec:pixels}.}

Next, we explain how we define each of the environments \mod{included in the masks}: bars (Sect.\,\ref{Sec:bars}), spiral arms (Sect.\,\ref{Sec:spirals}), rings (Sect.\,\ref{Sec:rings}), lenses (Sect.\,\ref{Sec:lenses}), bulges (Sect.\,\ref{Sec:bulges}), centres (Sect.\,\ref{Sec:centers}) and discs (Sect.\,\ref{Sec:discs}). Fig.~\ref{fig:intro} shows visual examples of these structures. The interarm regions are defined to be complementary to the spiral masks at the same galactocentric radii. Since a given pixel in the masks can belong to several morphological components simultaneously (e.g.\ a pixel in a nuclear ring can be on top of a bar),  we also propose a simple way to uniquely assign pixels to \mod{a dominant} environment (Sect.\,\ref{Sec:pixels}).

\begin{figure*}[t]
\begin{center}
\includegraphics[width=0.7\textwidth]{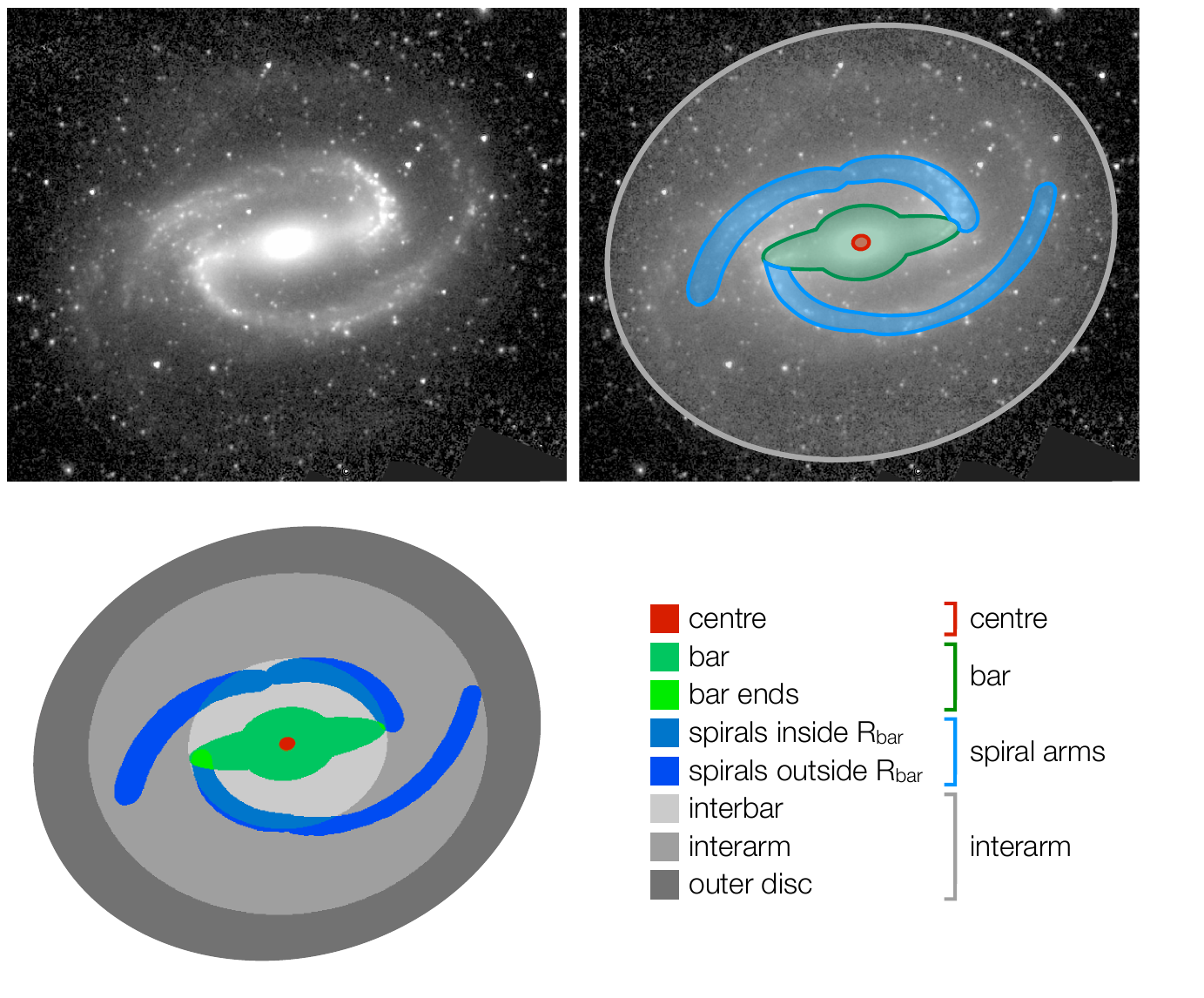}
\end{center}
\caption{Notation used in the `simple' masks, where each pixel is uniquely assigned to a dominant environment. The background image is the {\it Spitzer} $3.6$\,$\mu$m map of NGC\,1300 and the different colours denote different environments. Several of these environments can be grouped together for further simplicity, as indicated in the bottom-right diagram and in Table~\ref{table:masknotation}.
}
\label{fig:non-overlapping_masks}
\end{figure*}

\begin{table*}[t!]
\begin{center}
\caption[h!]{Notation in non-overlapping masks and simplified assignments.}
\begin{tabular}{ccccc}
\noalign{\smallskip}
\hline\hline
Label & Environment & & Label & Environment \\
   \hline
1 & centre (small bulge or nucleus)  & $\longrightarrow$ & 1 & centre \\
2 & bar (excluding bar ends) & \multirow{2}{*}{$\Big\}$} & \multirow{2}{*}{2$+$3} & \multirow{2}{*}{bar}  \\
3 & bar ends (overlap of bar and spiral) & & &  \\
5 & spiral arms inside interbar ($R_{\rm gal} < R_{\rm bar}$) & \multirow{2}{*}{$\Big\}$} & \multirow{2}{*}{5$+$6} & \multirow{2}{*}{spiral arms}  \\
6 & spiral arms ($R_{\rm gal} > R_{\rm bar}$) & & & \\ 
4 & interbar ($R_{\rm gal} < R_{\rm bar}$ but outside bar footprint) & \multirow{3}{*}{$\Bigg\}$} & \multirow{3}{*}{4$+$7$+$8} & \multirow{3}{*}{interarm}  \\
7 & interarm (only $R_{\rm gal}$ spanned by spiral arms, and $R_{\rm gal} > R_{\rm bar}$)  & & & \\ 
8 & outer disc ($R_{\rm gal} >$ spiral arm ends) in galaxies with spirals masks  & & & \\ 
9 & interbar ($R_{\rm gal} < R_{\rm bar}$) in galaxies without spiral masks & \multirow{2}{*}{$\Big\}$} & \multirow{2}{*}{9$+$10} & \multirow{2}{*}{disc without spiral masks} \\
10 & disc ($R_{\rm gal} > R_{\rm bar}$) in galaxies without spiral masks &  &  &  \\
  \hline
\end{tabular}
\label{table:masknotation}
\end{center}
\end{table*}

\subsection{Bars} 
\label{Sec:bars}

We define the contour enclosing each stellar bar as an ellipse. For a fixed centre, the bar ellipse is defined by three parameters: semi-major axis (bar size), axis ratio and position angle. We quote all of these parameters in the plane of the sky, without any deprojections. Most bars \mod{in PHANGS} have a projected half-length of a few kpc (with a median of $\sim$3\,kpc); three PHANGS galaxies are in fact double-barred systems, where we also implemented a nuclear bar.

For galaxies in S$^4$G, we mostly follow \citet[][\citetalias{2015A&A...582A..86H} hereafter]{2015A&A...582A..86H}, who homogeneously defined the size of bars visually on {\it Spitzer} $3.6$\,$\mu$m images. The visual bar lengths are in good agreement with automated methods to detect bars \citep{2013ApJ...771...59M}, but they tend to be more robust particularly when the emission associated with the bar is faint. Ellipse fitting was performed to measure the ellipticity of the bar given its length and position angle (PA) defined visually (by carefully varying the contrast of the images). In a few cases, it was not possible to determine the ellipticity via ellipse fitting in \citetalias{2015A&A...582A..86H}, and for these few cases we estimated the bar ellipticity visually. For galaxies outside S$^4$G, we relied on NIR measurements from the literature whenever possible, mostly from \citet{2007ApJ...657..790M}. We inspected the bars individually and when the bar seemed questionable (typically small bar candidates with less regular isophotes), we also examined optical images and kinematic information available from PHANGS to decide if a bar should be included or not.

For S$^4$G galaxies, we performed some minor modifications on the bar catalogue of \citetalias{2015A&A...582A..86H}. By default, we included the bars with good quality flags (1~or~2), excluding bars with quality flag~3. Some bars with the intermediate quality flag~2 seemed less reliable than others when considering our multi-wavelength data, and we decided to exclude bars in the following galaxies: NGC\,1385, NGC\,4424, NGC\,5042 and IC\,5332. There is a single target, NGC\,4941, where no bar was included in \citetalias{2015A&A...582A..86H}, but we decided to include it in the masks. This bar is in the catalogue of \citet{2007ApJ...657..790M}, and the photometric and kinematic information from PHANGS \citep{2020ApJ...897..122L} supports its existence. For NGC\,5248, we included the large-scale bar from \citet{2002ApJ...575..156J} instead of the smaller bar from \citetalias{2015A&A...582A..86H}. This is because our CO residual velocity map \citep{2020ApJ...897..122L} is suggestive of such a large-scale bar, but we note that other classifications did not include it \citep[e.g.][]{2015ApJS..217...32B}.

Outside S$^4$G, we adopted the bars from the catalogue in \citet{2007ApJ...657..790M}. We also included bars, with new visual measurements on {\it Spitzer} $3.6$\,$\mu$m images, for NGC\,2566 and NGC\,3059. Questionable cases include NGC\,2090, NGC\,2997 and NGC\,3137, but the multi-wavelength PHANGS information does not clearly support the existence of a bar, so we decided not to implement a bar for these galaxies.

Generally, we assumed that the centre of the bar matches the galaxy centre, which we adopted from Table~1 from \citet{2015ApJS..219....4S} for S$^4$G galaxies and from the PHANGS sample table \mod{version 1.6} for the remaining galaxies \citep{2021arXiv210407739L}. 
There are two galaxies where we adopted a different centre, as the bar seems clearly offset: IC\,1954 ($\rm RA=+52.880407$\,deg, $\rm Dec=-51.904783$\,deg) and NGC\,1559 ($\rm RA=+64.398638$\,deg, $\rm Dec=-62.783728$\,deg), offset by $1.6\arcsec$ and $6.3\arcsec$, respectively. We also included the two nuclear bars present in the sub-sample of PHANGS in \citetalias{2015A&A...582A..86H} (NGC\,1433 and NGC\,4321), and, outside S$^4$G, the nuclear bar in NGC\,1317 \citep{2004A&A...415..941E}.

\subsection{Spiral arms and interarm regions} 
\label{Sec:spirals}

Our goal is to define the 2D shape of strong spiral arms when they are dominant features across the galaxy disc. This is carried out as a three-step process: first, a log-spiral function is fitted to regions with bright $3.6$\,$\mu$m emission along each arm; second, these ideal log-spiral curves are assigned a width determined empirically; and, third, the starting and ending azimuth of each spiral segment is adjusted by eye to match the observations. The first step follows \citetalias{2015A&A...582A..86H}, and indeed we rely on their results for the vast majority of the S$^4$G galaxies. For galaxies outside S$^4$G, and for a few S$^4$G galaxies, new fits were performed following the same approach.

In detail, to perform the log-spiral fits, we first constructed a set of unsharp-masked versions of the {\it Spitzer} $3.6$\,$\mu$m images, which highlight spiral features. On the optimal unsharp-masked image, where the contrast is clearest, bright points along the arms were visually identified and their coordinates were recorded. These coordinates were deprojected to the plane of the galaxy (using the inclination and PA from the PHANGS sample table; \citealt{2020ApJ...897..122L}), and linear fits were performed for each segment in logarithmic polar coordinates. The results of these log-spiral fits were projected back to the plane of the sky. 

The strategy of assigning a finite width to the analytic log-spiral segments is new. The widths of spiral arms vary depending on the chosen tracer and resolution, and angular offsets are possible from one tracer to another (see e.g.\ \citealt{2013ApJ...779...42S,2016ApJ...827..103K,2017ApJ...845...78C,2017MNRAS.465..460E,2017ApJ...836...62S}). As our main goal is to capture the CO emission in spiral arms, we iteratively dilated the analytic spiral curves until a certain empirical threshold was reached. The dilated width in the plane of the galaxy was varied in multiples of half a kpc ($500$\,pc, $1000$\,pc, $1500$\,pc, $2000$\,pc, etc.). For each of these masks, we measured the total CO flux within the mask on the ALMA 7\,m+TP zeroth-order `broad' moment map (at ${\sim}7\arcsec$ resolution) and obtained the ratio with respect to the flux in the previous mask (i.e.\ $F_{\rm 1000\,pc} / F_{\rm 500\,pc}$, $F_{\rm 1500\,pc} / F_{\rm 1000\,pc}$, etc.). For this purpose, we used ALMA 7\,m+TP observations (instead of the higher-resolution 12\,m+7\,m+TP data) because spiral arms are more clearly identifiable and less subject to local irregularities. The final width was established when the ratio of CO flux from one step to the next falls below an empirical threshold of $1.25$ (i.e.\ the flux increases by less than $25$\% from one step to the next). This empirical convergence criterion results in spiral masks which are typically ${\sim}1{-}2$\,kpc wide and capture most of the $3.6$\,$\mu$m, CO and H$\alpha$ emission that one would associate with the arm by eye.

The above criterion is not always robust when a given spiral segment has limited coverage by ALMA. For spiral segments where ${<}30$\% of the pixels in the $3.6$\,$\mu$m spiral footprint have CO detections from ALMA 7\,m+TP, we did not use the above criterion and instead assigned a representative width as the mode of the remaining spiral segments in that galaxy. In the absence of any spiral segments with ${\geq}30$\% ALMA coverage for a given galaxy, the median ($1.5$\,kpc) width across the whole PHANGS sample was assigned.

Finally, we visually adjusted some of the endpoints of the spiral segments (i.e.\ the start and finish azimuth of the log-spiral) in order to provide continuity along coherent arms when there is a change in pitch angle (two or more segments trace a given continuous arm); this means that some segments which appeared to have a gap were merged to form a single spiral arm. \mod{This situation is illustrated by the northern spiral arm of NGC\,1300 shown in Fig.~\ref{fig:non-overlapping_masks}, where the pitch angle of the spiral arm changes.} By adjusting the endpoints of the segments, we also ensure that we correctly capture the spiral structures in the different tracers towards the inner and outer edge of the spiral arm.

We note that the catalogue from \citetalias{2015A&A...582A..86H} includes many spiral segments that we did not implement in our masks, including short and relatively isolated segments in multi-armed spirals. The difference between spiral arms and interarm regions becomes increasingly subjective as we move towards smaller and less continuous segments in multi-armed and flocculent spirals. This is why we conservatively defined spiral masks only in the cases where one can clearly trace a continuous and large-scale spiral structure covering a significant fraction of the disc \mod{(from the nucleus or end of the bar out to ${\gtrsim}1.5 R_{\rm e}$)}. Often this agrees with the cases where the galaxy has been defined as grand design. For example, for S$^4$G galaxies, $62$\% of the galaxies where we implemented spiral arms were defined as `grand design' by \citet{2015ApJS..217...32B}, while the remaining ones are mostly multi-armed \mod{(see e.g.\ NGC\,1637 and NGC\,4254 in Fig.~\ref{fig:atlas}, which were classified as multi-armed by \citealt{2015ApJS..217...32B} but have a mask with three spiral arms that span a large radial range)}.
Conversely, for $35$\% of the S$^4$G sub-sample classified as grand-design, no spiral arms were implemented in our masks; this can be due to their high inclination or because it was difficult to robustly define a binary spiral mask for other reasons. We optimised and verified our final spiral arms by inspecting the masks on molecular gas and star formation maps. 
This involved several rounds of quality flagging by three co-authors (MQ, ES, SEM), deciding segment by segment whether to keep it or not, adjusting the start or finish azimuth, and, in a few cases, modifying the width.

\subsection{Rings} 
\label{Sec:rings}

We also implemented rings for S$^4$G galaxies based on the visual measurements from \citetalias{2015A&A...582A..86H}. As shown by \citetalias{2015A&A...582A..86H}, the ring sizes show good agreement with previous identifications from the Near-InfraRed S0 Survey (NIRS0S) at $2.2$\,$\mu$m \citep{2011MNRAS.418.1452L} and the Atlas of Resonance (pseudo)Rings As Known In the S$^4$G (ARRAKIS) at $3.6$\,$\mu$m \citep{2014A&A...562A.121C},
in spite of small methodological differences (for example, \citetalias{2015A&A...582A..86H} used unsharp-masked images while \citet{2014A&A...562A.121C} relied on residual images, subtracting a GALFIT model).

We only keep the rings with best quality flags from the catalogue of \citetalias{2015A&A...582A..86H} (${\rm flag}=1$ or~2), avoiding pseudo-rings altogether, \mod{which are incomplete rings usually made of spiral arms}. \citetalias{2015A&A...582A..86H} did not determine the ring width, so we defined it visually on {\it Spitzer} $3.6$\,$\mu$m images.
The resulting width based on $3.6$\,$\mu$m typically covers well the extent of CO emission associated with the ring. We removed the nuclear ring in NGC\,4536, as \mod{the IRAC image shows a slight central flux depression, but not a real \mod{inner} boundary due to the smoothing imposed by the PSF.}

For galaxies outside S$^4$G, we performed our own measurements of the size and orientation of the rings on {\it Spitzer} $3.6$\,$\mu$m images, similarly to \citetalias{2015A&A...582A..86H}. 
Ring-lenses are intermediate structures between rings and lenses \citep[e.g.][]{2013A&A...555L...4C}; for practical purposes, we also delimited the two nuclear ring-lenses present in our sample (NGC\,1300 and NGC\,1433) as the ring that is visible in CO emission, defining the width based on the CO maps.

\subsection{Lenses} 
\label{Sec:lenses}

Lenses are morphological structures with relatively flat brightness profiles and a characteristic, well-defined outer edge, where the surface brightness falls off rapidly. They were defined as early as \citet{1961hag..book.....S} and \citet{1979ApJ...233..539K}, but only recently have they been systematically classified for large samples of galaxies. \citetalias{2015A&A...582A..86H} identified lenses across the S$^4$G sample, visually marking points along the edges of the lens and subsequently fitting them with an ellipse, providing the size, ellipticity and orientation of the structure. We only retain the lenses with best quality flag for our masks (${\rm flag}=1$ or~2). For galaxies outside S$^4$G, we measured the size and orientation of the most clear lenses similarly to \citetalias{2015A&A...582A..86H} on {\it Spitzer} $3.6$\,$\mu$m images (NGC\,2566, NGC\,5643, NGC\,6300).

It is worth distinguishing \textit{barlenses}, which are lens-like structures embedded in bars (typically spanning ${\sim}50$\% of the bar length). Their surface brightness drops fast at the edges and, unlike other lenses, they are thought to be the face-on counterparts of the vertically thick boxy/peanut structures of bars \citep{2011MNRAS.418.1452L,2014MNRAS.444L..80L,2015MNRAS.454.3843A}. In order to develop, barlenses seem to require a steep inner rotation curve which can originate from a central mass concentration such as a nuclear bulge \citep{2017A&A...598A..10L,2017ApJ...835..252S}.

Therefore, for most purposes, barlenses can be considered part of the bar and should be masked together. The full version of the masks includes barlenses within the `lens' category, together with lenses that are not associated with bars. Yet, in the simplified version of the masks, where pixels are uniquely assigned to a dominant environment (Sect.\,\ref{Sec:pixels}), barlenses are simply included as part of the bar footprint. The following PHANGS galaxies have a barlens (corresponding to the first lens in the masks, when more than one exists): NGC\,1097, NGC\,1300, NGC\,1512, NGC\,2566, NGC\,3351, NGC\,4548, NGC\,4579, NGC\,5134, NGC\,5643, and NGC\,6300.

\subsection{Bulges} 
\label{Sec:bulges}

Traditional photometric decompositions of galaxies usually distinguish an exponential disc from a central component named \textit{bulge} that is described by a certain S\'ersic index ($n$). The S$^4$G Pipeline~4 \citep{2015ApJS..219....4S} carried out systematic decompositions of the entire S$^4$G sample with GALFIT \citep{2002AJ....124..266P,2010AJ....139.2097P}, including up to two discs (to account for possible breaks), bulge, bar and an unresolved nuclear point source where applicable; they also used special functional forms for edge-on galaxies. It is important to emphasise that, since these GALFIT decompositions did not explicitly account for structures such as nuclear rings, the bulge component often contains rings and other nuclear structures. In this sense, some of the bulges are redundant with rings or lenses identified in \citetalias{2015A&A...582A..86H}. However, sometimes it is useful to have a generic \textit{bulge} flag, to permit simple comparisons without necessarily analysing detailed inner structures. Therefore, we keep bulges even when they clearly overlap with other structures, and allow the user to choose one component or the other depending on the specific goal.

In any case, we warn the reader that many of these `bulges' are likely not dispersion-dominated spheroids and might not even be `bulging' out of the main disc; that is why in this paper we prefer to group small bulges together with unresolved nuclear components under the deliberately more generic label of `centres' (see Sect.~\ref{Sec:centers}). Unlike bars, rings, or lenses, bulges do not have a well-defined morphological outer edge. Therefore, we arbitrarily define the edge of the bulge mask as twice the effective bulge radius. 

For galaxies outside S$^4$G, we adopt similar bulge measurements from the literature whenever possible. For example, we follow \citet{2004MNRAS.355.1251L}, who performed bulge-disc photometric decompositions on the $H$-band, for the following galaxies: NGC\,1317, NGC\,2090, NGC\,2566, NGC\,5643, and NGC\,6300.

\subsection{Centres} 
\label{Sec:centers}

The centre mask captures unresolved or marginally resolved stellar structures that are centrally concentrated ($R \lesssim 10\arcsec$, and most often $R \lesssim 5\arcsec$). A number of stellar structures can result in an excess of light on the IRAC images near the nucleus: an unresolved nuclear bar, a nuclear ring, or a nuclear disc, for example. Most of the galaxies with a bright nuclear PSF in IRAC are flagged as \textit{nucleus} in the S$^4$G Pipeline~4 \citep{2015ApJS..219....4S}. The bright nuclear region is often larger than the $1.7\arcsec$ FWHM of the IRAC PSF at $3.6$\,$\mu$m, but still cannot be resolved, for example, as a nuclear ring (nuclear rings have typical sizes of a few $100$\,pc, which makes the smaller rings hard or impossible to resolve for our most distant targets). 
For galaxies with unresolved central stellar structures, we visually inspected the data and defined a \textit{centre} mask that covers the area with a $3.6$\,$\mu$m and/or CO excess. In some cases, the centre appears axisymmetric, independent from the orientation of the galaxy; this is what we would expect, for example, if a very compact nuclear structure is broadened by the IRAC PSF. In other cases, however, the central structure appears elongated, usually with a similar PA and axis ratio as the main disc; this is expected if the structure is slightly extended and axisymmetric in the plane of the galaxy. Therefore, in each case we define the PA and axis ratio of the centre, in addition to its size, in order to accommodate these different possibilities. These nuclear masks are typically a few hundred parsec in radius (median $R \sim 300$\,pc), corresponding to $2{-}5$\% of $R_{25}$ with a few outliers.

Often we do not know what exact structure (or combination of structures) is causing the excess of light in the central few hundred parsecs, and this is why we adopt an agnostic approach by calling this `centre', alluding to the location rather than a physical component. A similar central excess of NIR flux is captured by small bulges identified with GALFIT (Sect.~\ref{Sec:bulges}); for many purposes, it makes sense to combine these small bulges with the `centres' defined here (Sect.~\ref{Sec:pixels}). This `centre' masks likely reflect regions similar to the central molecular zone (CMZ) in our galaxy, which corresponds to a nuclear stellar disc \citep{2002A&A...384..112L}.

\subsection{Discs} 
\label{Sec:discs}

For completeness, our masks include the discs identified via photometric decompositions of NIR images. These are the outer boundary of all galactic discs, not just what we call `discs without spirals' in this paper; the latter are simply the discs where we did not explicitly define spiral arms. For S$^4$G, we rely on the results from \citet{2015ApJS..219....4S}, which in a few cases include two different exponential discs within a given galaxy, implying a photometric break in the disc, and we also implemented those in our masks. Similarly to bulges, the edge of the disc was defined as twice the effective radius (an arbitrary choice, but visually reasonable). 
For galaxies outside S$^4$G the disc definitions were incorporated from the literature \citep[][]{2004MNRAS.355.1251L} or from new GALFIT decompositions of the {\it Spitzer} IRAC images. 
\mod{The disc mask does not play a role in the current paper, because we perform the analysis inside the ALMA field of view, which is smaller than this disc definition and therefore more restrictive. However, the disc component could be useful for future applications.}

\subsection{Uniquely assigning pixels to a dominant stellar structure} 
\label{Sec:pixels}

In our environmental masks, a given pixel can be assigned to multiple components. For instance, a pixel in a nuclear ring is also typically part of a bar. For some scientific applications, this multiple identity of a given position is useful, but in other cases it is more convenient to uniquely assign each pixel to a dominant stellar structure, as explained next. \mod{We publicly release both the full environmental masks and this simpler version where each pixel corresponds to a single environment.}

At the simplest level, we define five basic dominant environments: (1) centre, (2) bar, (3) spiral arms, (4) interarm, and (5) disc without spiral masks. This is illustrated in Fig.~\ref{fig:non-overlapping_masks} \mod{and summarised in Table~\ref{table:masknotation}}. To construct these simple masks, first we gave precedence to centres over any other components. Next, we labelled any remaining pixels within the bar footprint (if present) as bar. For galaxies with spiral masks, we separated the remaining area into spiral arms and interarm regions. Many galaxies do not have spiral masks (e.g.\ flocculent discs) and, in these cases, all the pixels beyond the centre or bar were flagged generically as `disc'.

We note that the `centre' environment in these simple masks includes both small bulges ($R \lesssim 10\arcsec$) that come out of bulge-disc photometric decompositions \citep{2004MNRAS.355.1251L,2015ApJS..219....4S}, and the `centre' flag that we introduced in Sect.~\ref{Sec:centers}. The `centre' often encompasses a nuclear ring, which is explicitly defined in the detailed masks, but embedded within the `centre' ellipse here; as mentioned above, these `centres' likely map structures comparable to the CMZ. These simple masks also consider barlenses as indistinguishable from the bar footprint (see Sect.~\ref{Sec:lenses} above for details). Fig.~\ref{fig:non-overlapping_masks} highlights the combined bar and barlens structure in NGC\,1300 in green.

Optionally, these non-intersecting masks also allow for slightly more sophisticated indexing of some environments, as shown in the bottom-left panel of Fig.~\ref{fig:non-overlapping_masks}. The tips of bars often overlap with the beginning of the spiral arms, and the overlapping pixels are assigned a different label (`bar ends'). The masks also permit to isolate the interarm regions that fall within the bar radius, which some authors prefer to call `interbar' \citep[e.g.][]{1985ApJ...288..438E,2009ApJ...703.1297E}. For completeness, the spiral arms that are inside the bar radius but outside the bar footprint, which often arch to form pseudo-rings, are also given a different code. Finally, when spiral arms do not reach out to the end of the disc, the area beyond the end of spiral masks is differentiated (`outer disc').

It is up to the user whether to merge some of these environments together, depending on the level of detail required \mod{(see Table~\ref{table:masknotation})}. For this paper, we uniquely assign pixels to a dominant stellar structure and focus on the five simple environments defined in this section (centre, bar, spiral arms, interarm regions, and discs without spiral masks), as in the top-right panel of Fig.~\ref{fig:non-overlapping_masks}.

\section{Results} 
\label{Sec:results}

\begin{figure*}[t]
\begin{center}
\includegraphics[trim=0 160 0 0, clip,width=0.8\textwidth]{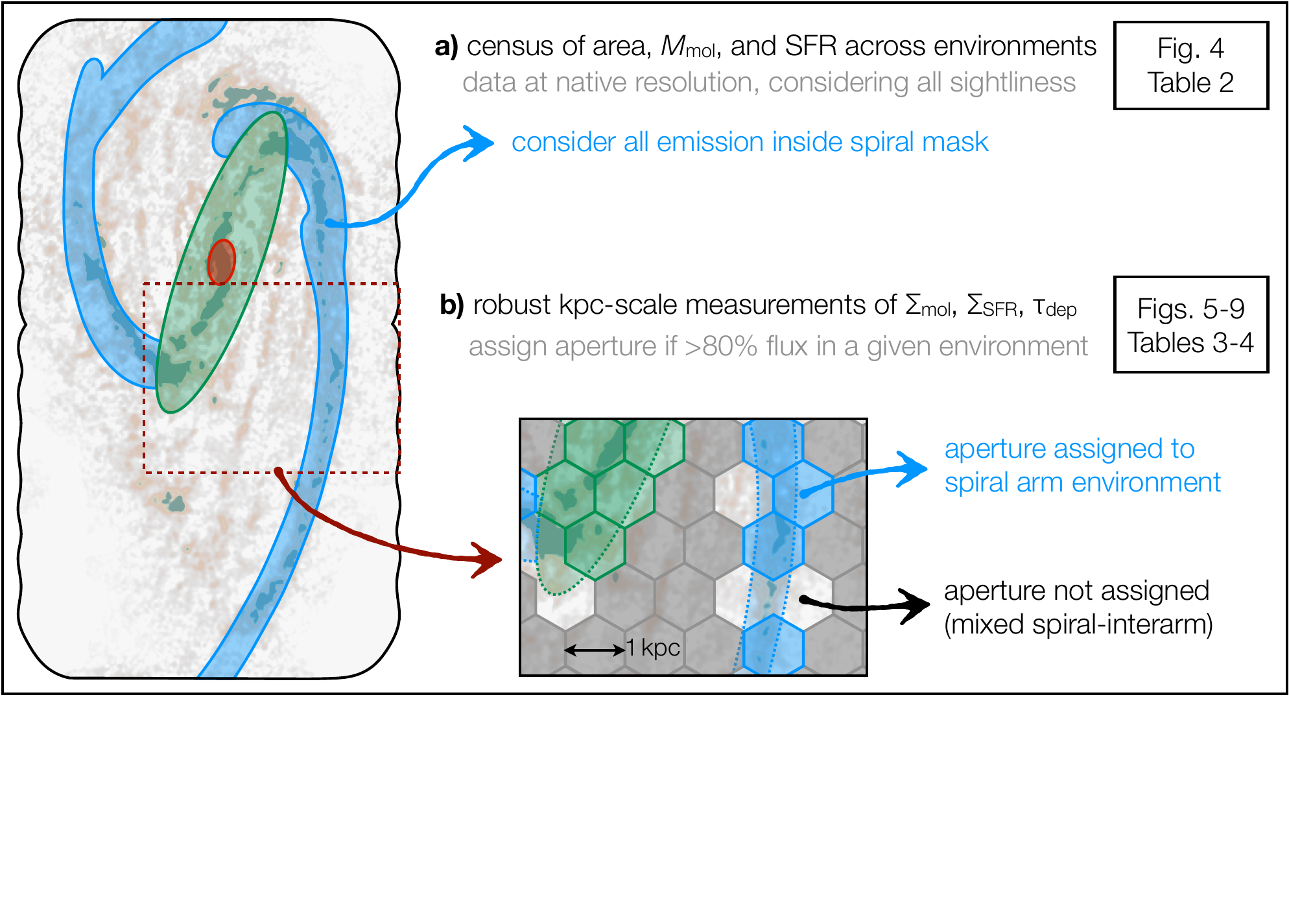}
\end{center}
\caption{\mod{Strategy followed to assign emission to environments illustrated with NGC\,3627. For the census of area, $M_{\rm mol}$, and SFR as presented in Sect.~\ref{Sec:piechart}, we consider all the emission within the mask footprint of each environment (working at our native resolution of ${\sim}1\arcsec$). For the analysis presented in Sects.~\ref{Sec:surfdens}--\ref{Sec:contrasts}, we turn to measurements on kpc-sized hexagonal apertures, so that we implicitly average over the star-forming cycle and SFRs are more robustly estimated.}
}
\label{fig:strategy}
\end{figure*}

\subsection{Census of molecular gas and star formation across environments}
\label{Sec:piechart}

\begin{figure}[t]
\begin{center}
\includegraphics[trim=0 0 0 0, clip,width=0.47\textwidth]{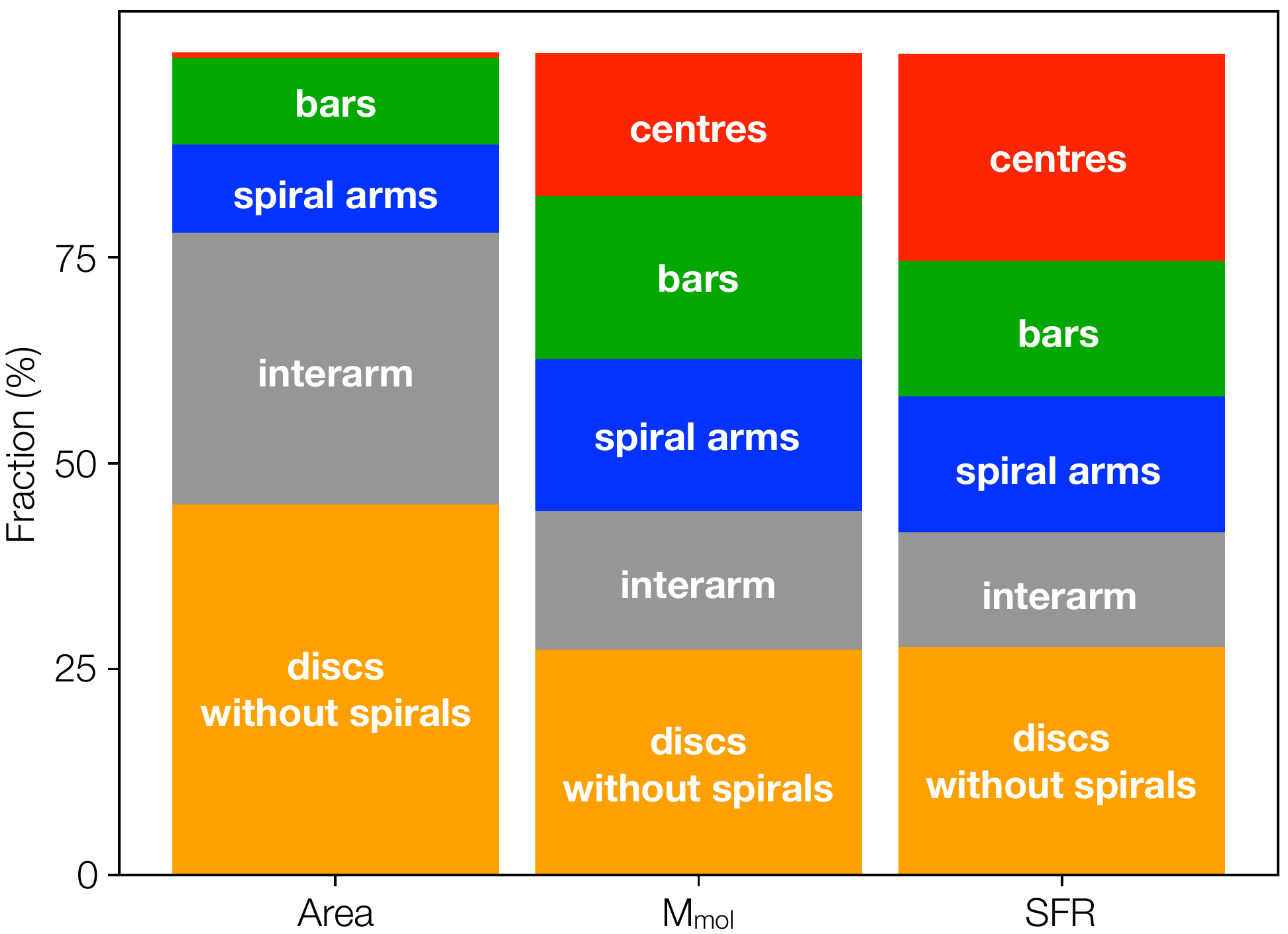}
\end{center}
\caption{Stacked bar charts showing the relative distribution of area, integrated molecular gas mass, and integrated star formation rates in each of the environments that we consider in this paper and across the entire PHANGS--ALMA sample of galaxies. These measurements consider the full resolution of the data as explained in Sect.~\ref{Sec:piechart} and are limited to the ALMA field of view.
}
\label{fig:pie_charts}
\end{figure}

\begin{table*}[t!]
\begin{center}
\caption[h!]{Distribution of area, molecular gas mass and SFRs across environments.}
\begin{tabular}{lcccccc}
\hline\hline
\noalign{\smallskip}
 & centre & bar & spiral & interarm & disc & all  \\
\noalign{\smallskip}
   \hline
\noalign{\smallskip}
Area [kpc$^2$] & 92 (0.66\,\%) & 1467 (10.5\,\%) & 1495 (10.7\,\%) & 4620 (33.0\,\%) & 6296 (45.0\,\%) & 13972 (100\,\%) \\
H$_2$ mass [$10^{10}$\,$M_\odot$] & 2.32 (17.4\,\%) & 2.64 (19.8\,\%) & 2.45 (18.4\,\%) & 2.25 (16.8\,\%) & 3.66 (27.4\,\%) & 13.3 (100\,\%) \\
SFR [$M_\odot$\,yr$^{-1}$] & 23.9 (25.2\,\%) & 15.6 (16.4\,\%) & 15.6 (16.5\,\%) & 13.1 (13.9\,\%) & 26.2 (27.7\,\%) & 94.6 (100\,\%) \\
\noalign{\smallskip}
 \hline
 \hline
\end{tabular}
\label{table:stats}
\end{center}
\tablefoot{\mod{These measurements integrate} the area or flux across each environment for all PHANGS galaxies at the highest resolution available to us (${\sim}1\arcsec$) within the ALMA field of view. As explained in the text, the SFR budget is dominated by a few outliers, where AGN contamination might also be an issue; excluding the four most star-forming galaxies, centres would contribute $11$\% of the total SFR instead of $25.2$\%.}
\end{table*}

We start by examining how the integrated molecular gas mass, star formation rates and area are distributed among the environments that we consider in this paper: centre, bar, spiral arms, interarm, and discs without spiral masks. Not all galaxies have each of these environments, but we consider the relative contribution of these regions to the total budget of area, molecular gas and star formation across the $74$ galaxies in the main PHANGS--ALMA sample. All measurements were restricted to the ALMA field of view and they were performed at the highest resolution available to us: the native resolution from ALMA for the molecular gas and the narrow-band H$\alpha$ maps described in Sect.~\ref{Sec:SFRs} for the SFRs. 
\mod{As explained in Sect.~\ref{Sec:PHANGS--ALMA}, the ALMA field of view represents the area where we expect to find most molecular gas and star formation.}
\mod{Fig.~\ref{fig:strategy} illustrates the strategy followed in this subsection (contrasting it to the approach from the following subsections), and} Table~\ref{table:stats} lists these measurements.

The stacked bar charts in Fig.~\ref{fig:pie_charts} show how the integrated molecular gas mass, SFRs and areas are split into environments. We limit this comparison to pixels inside the ALMA field of view, and consider the area in the plane of the galaxies (i.e.\ corrected for inclination).
The main contribution in terms of total area across PHANGS comes from the interarm environment, closely followed by discs without spiral masks (mostly multi-armed and flocculent spirals); when added up, these two environments roughly cover $75$\% of the area across PHANGS galaxies. Spiral arms, bars, and centres make up less than one quarter of the total area spanned by PHANGS--ALMA. However, when we focus on the contribution to the molecular mass or star formation, the view is quite different.
Molecular gas mass is quite evenly distributed among the environments that we consider. Centres cover a very small area ($0.7$\%), but their contribution to the global molecular mass and star formation budget is remarkably high.
The contribution of centres to total SFR ($26$\%) is higher than their contribution to molecular gas mass ($18$\%). This is largely driven by a few starburst galaxies, where the central SFR can exceed the total SFR from other entire galaxies \citep[e.g.][]{1998ApJ...498..541K}. If we exclude the four most extreme central star-forming galaxies (NGC\,1365, NGC\,1672, NGC\,2566 and NGC\,7496), in some of which AGN contamination might also be an issue, the contribution of centres to the star formation budget across the whole sample drops \mod{from $25.2$\% to $11$\%. However, if we exclude those starburst galaxies, the molecular gas mass in centres only drops from $17.4$ to $11$\%, while the area decreases from $0.66$ to $0.6$\%.} Apart from centres, the other four environments contribute a similar share of the integrated SFRs, in spite of the significantly different area they cover. This immediately points to the idea that the average surface densities in centres, bars, and spiral arms must be higher than in interarm regions or discs without spirals, as we show next.

In Appendix~\ref{sec:Appendix1} we list measurements analogous to Table~\ref{table:stats}, but restricted to the galaxies where ALMA covers a higher relative fraction of the galaxy discs (${\rm FoV} > R_{25}$).  \mod{This} confirms that the results in Table~\ref{table:stats} and Fig.~\ref{fig:pie_charts} are not strongly influenced by differences in the ALMA coverage among galaxies.

\subsection{Molecular gas and star formation rate surface density: radial trends}
\label{Sec:surfdens}

\begin{figure*}[t]
\begin{center}
\includegraphics[trim=0 0 0 0, clip,width=1.0\textwidth]{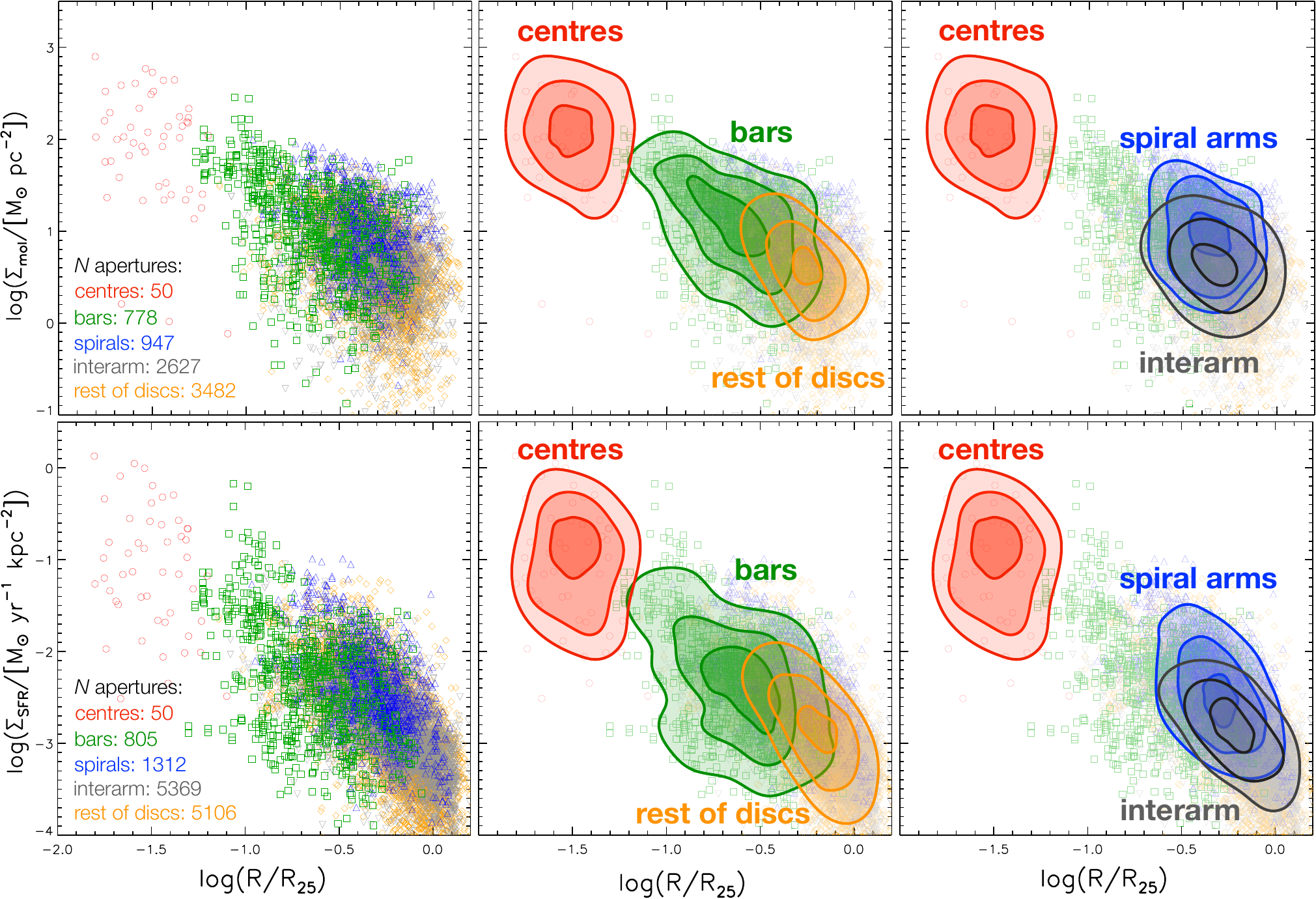}
\end{center}
\caption{Molecular gas and star formation rate surface densities as a function of galactocentric radius (normalised to $R_{25}$) for kpc-scale measurements across the PHANGS--ALMA sample of galaxies. The colours correspond to the different environments that we analyse in this paper. For clarity, the middle and right panels display alternative versions of the plots showing contours of data point density for different environments, highlighting their relative offsets \mod{(contours encompass 30\%, 50\%, and 80\% of all data points in each category)}.
}
\label{fig:SurfDens_vs_Rnorm}
\end{figure*}

\begin{figure*}[t]
\begin{center}
\includegraphics[trim=0 0 0 0, clip,width=1.0\textwidth]{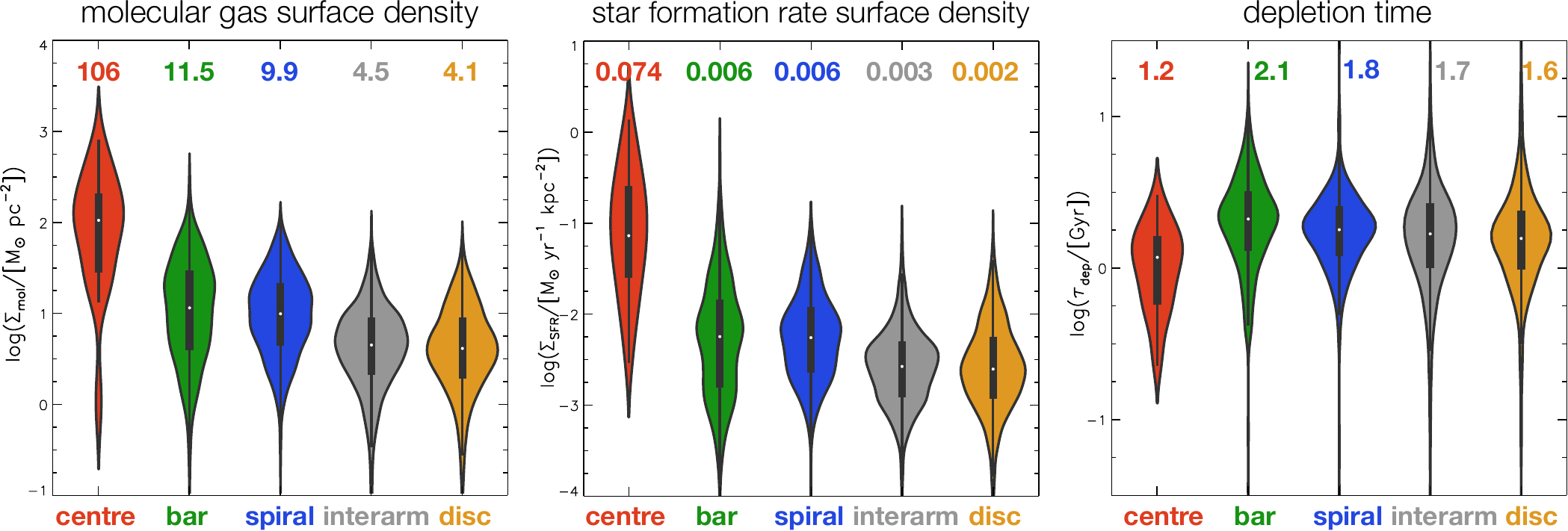}
\end{center}
\caption{Violin plots showing the distribution of molecular gas and star formation rate surface densities measured in 1\,kpc apertures, as well as the resulting depletion times ($\tau_{\rm dep} = \Sigma_{\rm mol}/\Sigma_{\rm SFR}$). The different colours indicate the range of environments that we examine in this paper. The numbers on top of the violin plots indicate the median value in linear scale. \mod{The thick black bar inside each violin plot shows the interquartile range, the white dot indicates the median, and the thin black lines show the span of data points beyond the black bar that lie within $1.5$ times the interquartile range.}
}
\label{fig:histo_surfdens}
\end{figure*}

\begin{table*}[t!]
\begin{center}
\caption[h!]{Distribution of molecular gas surface density, SFR surface density, and depletion times across environments.}
\begin{tabular}{llcccccc}
\hline\hline
\noalign{\smallskip}
 & & centre & bar & spiral & interarm & disc & all  \\
\noalign{\smallskip}
   \hline
   \noalign{\smallskip}
 \multirow{3}{*}{$\Sigma_\mathrm{mol}/(M_\odot~\mathrm{pc^{-2}})$ } &  median & $106.1_{-82.97}^{+225.5}$ & $11.51_{-8.620}^{+27.45}$ & $9.903_{-6.637}^{+16.91}$ & $4.492_{-2.924}^{+6.079}$ & $4.120_{-2.631}^{+7.610}$ & $5.001_{-3.290}^{+10.57}$ \\
 &  mean & $159.4$ & $20.92$ & $14.97$ & $6.589$ & $6.671$ & $10.03$ \\
 &  weighted & $340.8$ & $56.14$ & $29.82$ & $14.66$ & $15.15$ & $59.26$ \\
    \noalign{\smallskip}
  \hline
   \noalign{\smallskip}
 \multirow{3}{*}{$\Sigma_{\rm SFR}/(M_\odot~\mathrm{yr^{-1}}~\mathrm{kpc^{-2}})$ } &  median & $0.0739_{-0.0592}^{+0.3406}$ & $0.0057_{-0.0045}^{+0.0171}$ & $0.0055_{-0.0037}^{+0.0100}$ & $0.0026_{-0.0016}^{+0.0034}$ & $0.0025_{-0.0015}^{+0.0052}$ & $0.0029_{-0.0018}^{+0.0061}$ \\
 &  mean & $0.2079$ & $0.0156$ & $0.0094$ & $0.0042$ & $0.0046$ & $0.0074$ \\
 &  weighted & $0.4956$ & $0.0596$ & $0.0198$ & $0.0094$ & $0.0108$ & $0.0715$ \\
   \noalign{\smallskip}
   \hline
   \noalign{\smallskip}
\multirow{3}{*}{$\tau_{\rm dep}/\mathrm{(Gyr)}$} &  median & $1.177_{-0.652}^{+0.650}$ & $2.102_{-1.041}^{+1.637}$ & $1.788_{-0.735}^{+1.038}$ & $1.677_{-0.872}^{+1.369}$ & $1.565_{-0.746}^{+1.245}$ & $1.671_{-0.824}^{+1.309}$ \\
 &  mean & $1.188$ & $2.432$ & $2.007$ & $2.459$ & $2.237$ & $2.296$ \\
 &  weighted & $0.978$ & $2.238$ & $1.949$ & $2.528$ & $2.106$ & $2.083$ \\
\noalign{\smallskip}
\hline
 \hline
\end{tabular}
\label{table:stats_surfdens}
\end{center}
\tablefoot{The measurements were performed in hexagonal apertures at kpc scales. The median and mean correspond to unweighted measurements \mod{(the uncertainties list} the difference between the median and the 16th and 84th percentiles), while the weighted measurements show the CO-weighted averages.}
\end{table*}

Here we examine how local surface densities (of molecular gas and SFR) vary from environment to environment. 
Instead of using the high-resolution maps from Sect.\,\ref{Sec:piechart}, the subsequent analysis is performed at kpc scales, which allows for more robust estimates of star formation rates\mod{; this approach is illustrated in Fig.~\ref{fig:strategy}.}
\mod{We extracted measurements at the centre of each kpc-scale hexagonal aperture from surface density maps at a common physical resolution of $1.5$~kpc \citep[as explained in][]{2020ApJ...892..148S}.}
As described in \mod{Sect.~\ref{Sec:SFRs}}, the SFR estimates combine \textit{GALEX} FUV and \textit{WISE} $22$\,$\mu$m data to capture both the unobscured and obscured star formation; the conversion from CO to molecular gas mass relies on a radially varying prescription based on the estimated local metallicity. \mod{All surface densities are expressed in the plane of the galaxies, corrected for inclination.}

Each of the hexagonal apertures, with a separation of $1$\,kpc between adjacent centres, provides a data point to the plots that follow. A given hexagonal aperture is assigned to one of the environments considered here (centre, bar, spiral, interarm, or disc without spiral masks) if at least $80$\% of the CO emission and SFR in the hexagon falls within the footprint of a given environment \mod{(Fig.~\ref{fig:strategy})}. In this sense, the aperture designation is flux-weighted; for instance, a hexagonal aperture that overlaps 45\% in area with a spiral arm and 55\% with interarm is assigned to the spiral environment if 80\% of the emission it captures actually arises from the spiral arm (when measured using the high-resolution maps). In Appendix~\ref{sec:Appendix1} we confirm that varying this flux threshold between $70$\% and $90$\% does not strongly impact our results. For this assignment purpose we use the maps at native resolution in order to mitigate the potential dilution and redistribution of flux among environments at lower resolution. This means that not all 1\,kpc measurements are included in the plots; for instance, a hexagonal aperture where 60\% of the flux comes from a bar and 40\% comes from the interarm is not assigned to any environments. Specifically, the $80$\% flux threshold excludes $29$\% of the hexagonal apertures. This strategy ensures that we plot only the measurements that are reliably associated with a single environment and avoid those that are mixed. The following plots also consider the normalised radius for each aperture, \mod{normalised by $R_{25}$ (the semi-major axis of the $B$-band $25$\,mag\,arcsec$^{-2}$ isophote). For the centre masks, we use instead the CO-weighted mean galactocentric radius (in order to avoid $\log(R=0)$).}

Figure~\ref{fig:SurfDens_vs_Rnorm} clearly highlights how centres harbour the highest molecular gas and SFR surface densities in PHANGS. The range of observed SFRs in the centres is slightly larger than the dynamic range in molecular gas surface density, pointing to variations in SFE in centres. This is not surprising, as some centres are starbursts while others are relatively quiescent, in spite of having large molecular gas surface densities.

Beyond centres, there is some trend for lower gas surface densities at larger radii, particularly in bars. However, even at fixed normalised radius, surface densities span up to $2{-}3$~dex, which highlights the huge diversity within and among galaxies. In other words, molecular gas surface densities do not always scale analogously with radius, and can be strongly affected by local environmental conditions \citep[e.g.][]{2019A&A...625A..19Q}. The most obvious additional factor driving the spread in molecular gas and SFR surface density at fixed radius is stellar surface density. This quantity shows a strong correlation with both the molecular gas distribution and the SFR \citep[e.g.][]{2001ApJ...561..218R,2002ApJ...569..157W,2008AJ....136.2782L,2021MNRAS.503.3643B,2021A&A...650A.134P} that has been described as a pressure--molecular gas correlation \citep[e.g.][]{2002ApJ...569..157W,2006ApJ...650..933B,2008AJ....136.2782L}, an extended star formation relation \citep{1993ApJ...418..804D,2011ApJ...733...87S,2021MNRAS.503.1615S}, or a resolved star-forming main sequence \citep{2019ApJ...884L..33L}.

The clouds of points corresponding to spiral arms, interarm regions, and discs without spiral masks largely overlap in the plane of $\Sigma_{\rm CO}$ (or $\Sigma_{\rm SFR}$) versus normalised radius. Interestingly, the surface densities in spiral arms are comparable to those in bars, despite being located at larger radii. If we compare spiral arm and interarm environments, there is a subtle but noticeable bulk vertical offset, implying that spiral arms have on average higher molecular gas and SFR surface densities. The centroids of the innermost contours are displaced vertically by $0.3{-}0.4$\,dex, which results in median surface densities that are roughly a factor of two higher in spiral arms than in the interarm environment (see also Fig.~\ref{fig:histo_surfdens} below).
This was shown by \citet{2020ApJ...901L...8S} for PHANGS data on cloud-scales and we further quantify it at kpc scales in Sect.~\ref{Sec:contrasts}. In any case, the overlap between spirals and interarm is large, which means that we can very often find some surface densities in the interarm environment which are higher than other spiral arm surface densities at a given normalised radius \mod{\citep[see also][for a study of the radial variation of arm and interarm molecular gas surface densities in M51]{2013MNRAS.433.1837V}}. Even if we consider a fixed normalised radius, we are looking at a mix of sight lines from vastly different kinds of galaxies, which cover a wide range of stellar masses and Hubble types.

\subsection{Surface densities and depletion time across environments}

In Figure~\ref{fig:histo_surfdens} we use violin plots to visualise the distribution across environments of surface densities and depletion time ($\tau_{\rm dep} = \Sigma_{\rm mol}/\Sigma_{\rm SFR}$\mod{, which is the inverse of the star formation efficiency, ${\rm SFE} = 1/\tau_{\rm dep}$)}. We only consider measurements from apertures with simultaneous detections in $\Sigma_{\rm mol}$ and $\Sigma_{\rm SFR}$, to ensure that we are consistently comparing the same sets of sight lines in the three panels. Centres clearly stand out as the environments with the largest average molecular gas and SFR surface densities; bars and spirals tend to harbour slightly higher surface densities than other environments, but the differences become more subtle. In spite of the large variation in median surface density among environments, the range of median depletion times is quite small \mod{($1.18$ to $2.10$\,Gyr)}.
\mod{Centres have the shortest depletion times (i.e.\ they are more efficient at forming stars), with a median of $1.2$\,Gyr, while bars have the longest depletion times (median of $2.1$\,Gyr). The other environments have intermediate depletion times (median of ${\sim}1.6{-}1.8$\,Gyr).} The shorter depletion times found in centres are consistent with previous findings in M51 \citep{2017ApJ...846...71L}.

Table~\ref{table:stats_surfdens} lists the medians and scatter in the distributions shown in Fig.~\ref{fig:histo_surfdens}. Additionally, it also lists the means and the CO-weighted averages. The violin plots show the unweighted distributions, considering all sight lines equally (i.e.\ weighting by area), which tells us about the typical expectation if we look at a random location within each of these environments. Weighting the kpc-size apertures by their molecular gas content captures the properties and depletion times that we can expect if we randomly pick up a molecular cloud in each of these environments. The unweighted mean for the molecular gas and SFR surface density is always higher than the median, implying that the distributions are skewed towards high values; this is not surprising, as there are substantial local enhancements in the surface densities and this is also the expectation for gas in a lognormal distribution. The mean surface densities become even higher if we weight by CO, which is expected by construction for molecular gas, and indirectly for star formation, since it follows molecular gas to first order. Weighted by CO, the characteristic depletion times tend to be slightly longer.

\begin{figure*}[t]
\begin{center}
\includegraphics[trim=0 0 0 0, clip,width=0.95\textwidth]{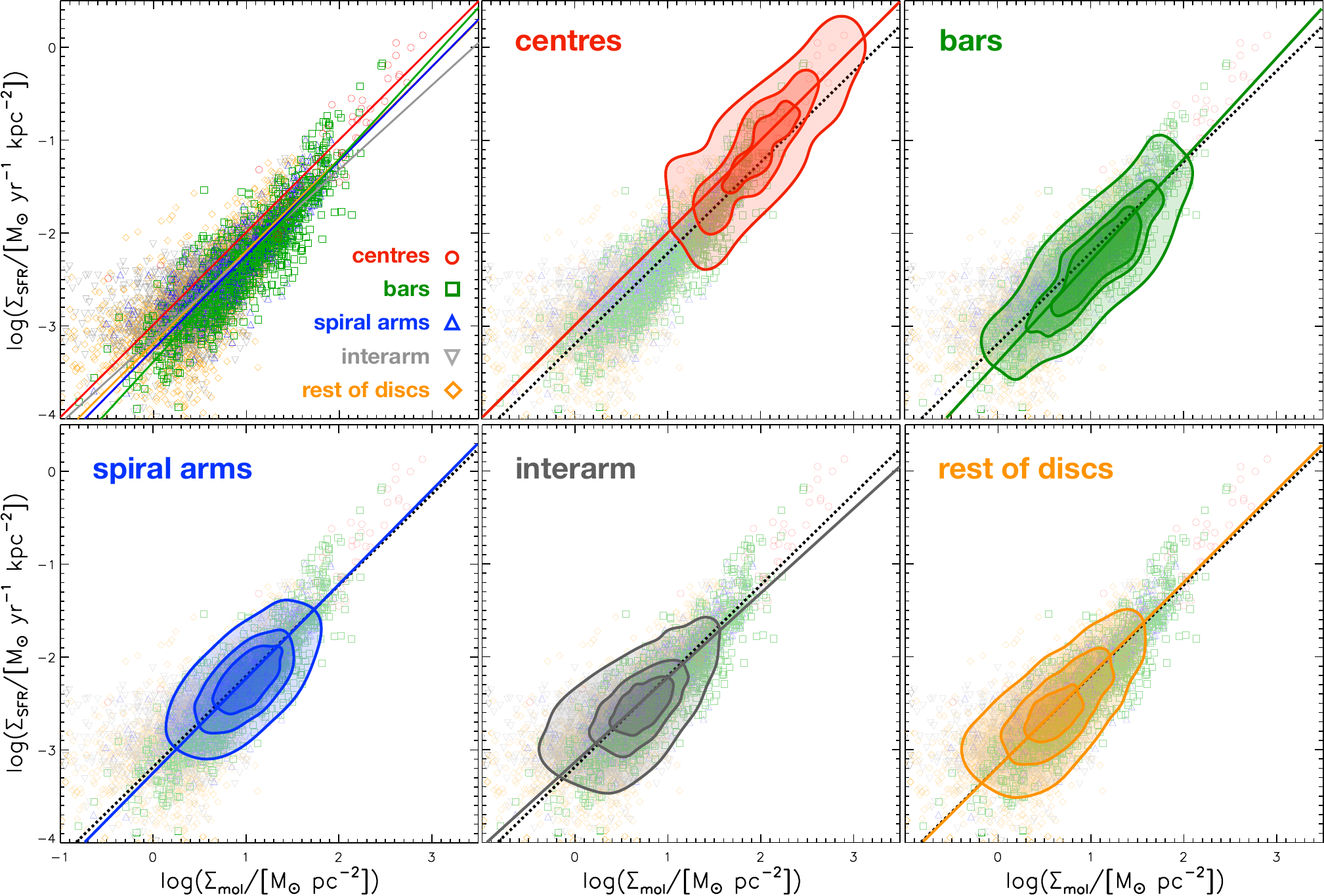}
\end{center}
\caption{Molecular Kennicutt--Schmidt relation for kpc-scale measurements across the PHANGS--ALMA sample of galaxies. The straight colour lines represent the best bisector fit to the data for each environment (for reference, the black dotted line represents the fit to all the data). For clarity, the various panels show the same plot with data point density contours for each environment and the corresponding bisector fit \mod{(contours encompass 30\%, 50\%, and 80\% of all data points in each category)}.
}
\label{fig:KSplot}
\end{figure*}

Finally, we examine the relation between molecular and SFR surface densities, known as the molecular Kennicutt--Schmidt relation \citep{1959ApJ...129..243S,1998ApJ...498..541K}. Fig.~\ref{fig:KSplot} shows this relation colour-coded by environment, and including the best-fit power-law regressions to the data using a bisector fit. The slopes and intercepts that we find for the different environments are listed in Table~\ref{table:KSfits}, relative to the same units as Fig.~\ref{fig:KSplot}, that is, $\log(\Sigma_{\rm SFR} /[M_\odot\,{\rm yr}^{-1}\,{\rm kpc}^{-2}]) = M + N \log(\Sigma_{\rm mol} /[M_\odot\,{\rm pc}^{-2}])$.
\mod{The slope for centres appears} steeper, but all environments have values consistent with each other within their uncertainties.
Table~\ref{table:KSfits} confirms that there are some differences in the slopes (and intercepts) depending on the adopted $\alpha_{\rm CO}$ prescription; the centre environment is most sensitive to the choice of $\alpha_{\rm CO}$. A constant Galactic conversion factor results in fairly similar slopes and intercepts, with differences of at most a few percent. Adopting the \citetalias{2013ARA&A..51..207B} prescription for $\alpha_{\rm CO}$ that explicitly depends on CO intensity \citep[see][for details]{2020ApJ...892..148S} yields larger departures, with a slope as high as $N=1.43$ (but also with a larger uncertainty) for all PHANGS sight lines. In the Appendix, Fig.~\ref{fig:KSplot_alphaCO} shows alternative plots to Fig.~\ref{fig:KSplot} using these different conversion factors.

In any case, for our preferred PHANGS $\alpha_{\rm CO}$ approach, the slopes for all of the environments are compatible with a linear relation within the uncertainties of the fits \mod{(within $1\sigma$, except for the interarm fit, where the offset is $1.6\sigma$)}. The slope fitted to all the PHANGS data points together is close to~1 ($N = 0.97 \pm 0.06$). This agrees with previous findings, where the molecular Kennicutt--Schmidt relation was found to be more linear than the atomic version; for instance, \citet{2008AJ....136.2846B} found $N=1.01$ from a combined analysis of a large number of sight lines from the HERACLES survey \citep{2009AJ....137.4670L}; \citet{2013AJ....146...19L} further refined these calculations, highlighting the role of the $\alpha_{\rm CO}$ conversion factor, and consistently find a slope of $N \approx 1.0$ in agreement with ours. Previous studies based on different surveys also recover an approximately linear relation between molecular gas and SFR surface density \citep[e.g.][]{2009ApJ...704..842B,2011AJ....142...37S,2017ApJ...846..159B,2017ApJ...846...71L,2019ApJ...872...16D,2019MNRAS.488.1926D,2019ApJ...884L..33L,2021MNRAS.501.4777E,2021arXiv210407739L}. \mod{Yet, we warn that the precise slope of the Kennicutt-Schmidt relation is sensitive to the masking and sampling scheme followed, as measurements at low signal-to-noise regions can affect the slope. Thus, comparisons among surveys must be done with caution.}

We also measure the median vertical offset of each environment with respect to the global fit to all data points in the Kennicutt--Schmidt plane. While $\tau_{\rm dep}$ variations may capture the key physical quantity, these offsets from an overall scaling remove any zeroth-order dependence of $\tau_{\rm dep}$ on gas surface density. The standard deviation in $\log(\Sigma_{\rm SFR})$ at fixed molecular gas surface density is fairly similar among environments ($0.24{-}0.35$\,dex), but there are bulk differences ranging from a median offset of \mod{$-0.11$\,dex for bars and spiral arms, up to $+0.23$\,dex for centres (interarm and discs without spiral masks have smaller offsets, $-0.06$ and $0.06$, respectively). This means that centres tend to have shorter depletion times (as seen in Fig.~\ref{fig:histo_surfdens}).}

\begin{table*}[t!]
\begin{center}
\caption[h!]{Bisector fits to the molecular Kennicutt--Schmidt relation for the different $\alpha_{\rm CO}$ prescriptions.}
\begin{tabular}{llcccc}
\hline\hline
 & & PHANGS $\alpha_{\rm CO}$ & Constant $\alpha_{\rm CO}$ & N12 $\alpha_{\rm CO}$ & B13 $\alpha_{\rm CO}$ \\
  \hline
\multirow{3}{*}{Centre} & slope & 1.00 $\pm$ 0.44 & 1.02 $\pm$ 0.50 & 1.71 $\pm$ 1.67 & 1.98 $\pm$ 2.51 \\
 & intercept & -2.99 $\pm$ 0.22 & -3.24 $\pm$ 0.20 & -4.31 $\pm$ 0.22 & -4.03 $\pm$ 0.21 \\
 & Spearman $\rho$ & 0.92 & 0.92 & 0.85 & 0.79 \\
 \hline
\multirow{3}{*}{Bar} & slope & 1.09 $\pm$ 0.17 & 1.04 $\pm$ 0.16 & 1.41 $\pm$ 0.38 & 1.87 $\pm$ 0.88 \\
 & intercept & -3.38 $\pm$ 0.04 & -3.43 $\pm$ 0.03 & -4.11 $\pm$ 0.05 & -4.49 $\pm$ 0.10 \\
 & Spearman $\rho$ & 0.89 & 0.88 & 0.88 & 0.78 \\
 \hline
\multirow{3}{*}{Spiral} & slope & 1.01 $\pm$ 0.18 & 0.95 $\pm$ 0.16 & 1.14 $\pm$ 0.27 & 1.42 $\pm$ 0.50 \\
 & intercept & -3.25 $\pm$ 0.03 & -3.19 $\pm$ 0.03 & -3.60 $\pm$ 0.06 & -3.90 $\pm$ 0.10 \\
 & Spearman $\rho$ & 0.89 & 0.89 & 0.89 & 0.83 \\
  \hline
\multirow{3}{*}{Interarm} & slope & 0.91 $\pm$ 0.10 & 0.82 $\pm$ 0.08 & 0.97 $\pm$ 0.19 & 1.26 $\pm$ 0.37 \\
 & intercept & -3.14 $\pm$ 0.02 & -3.07 $\pm$ 0.01 & -3.43 $\pm$ 0.03 & -3.84 $\pm$ 0.06 \\
 & Spearman $\rho$ & 0.74 & 0.76 & 0.81 & 0.75 \\
 \hline
\multirow{3}{*}{Disc} & slope & 0.99 $\pm$ 0.09 & 0.91 $\pm$ 0.08 & 1.08 $\pm$ 0.17 & 1.40 $\pm$ 0.42 \\
 & intercept & -3.19 $\pm$ 0.01 & -3.10 $\pm$ 0.01 & -3.50 $\pm$ 0.02 & -3.87 $\pm$ 0.04 \\
 & Spearman $\rho$ & 0.81 & 0.78 & 0.80 & 0.66 \\
 \hline
\multirow{3}{*}{All} & slope & 0.98 $\pm$ 0.05 & 0.89 $\pm$ 0.04 & 1.12 $\pm$ 0.10 & 1.47 $\pm$ 0.24 \\
 & intercept & -3.19 $\pm$ 0.01 & -3.12 $\pm$ 0.01 & -3.57 $\pm$ 0.02 & -3.99 $\pm$ 0.04 \\
 & Spearman $\rho$ & 0.83 & 0.81 & 0.83 & 0.74 \\
 \hline
\end{tabular}
\label{table:KSfits}
\end{center}
\tablefoot{The $p$-values of the Spearman rank correlation coefficients ($\rho$) shown in this table are very low (always ${<}1\%$), suggesting that the correlations are significant (except for centres with the B13 prescription, $p$-value$=$0.44).}
\end{table*}

\subsection{Surface density contrasts within a given galaxy}
\label{Sec:contrasts}

In the previous subsections we plotted together kpc-scale measurements arising from potentially very different galaxies, thus contrasting not only environments within a given galaxy, but also among galaxies. This highlights the diversity in molecular gas properties and its ability to form stars in analogous stellar structures in different galaxies. Now, we pose a related but slightly different question: `Within a given galaxy, do surface densities and depletion times differ among environments?'

\begin{figure*}[t]
\begin{center}
\includegraphics[trim=0 0 0 0, clip,width=0.65\textwidth]{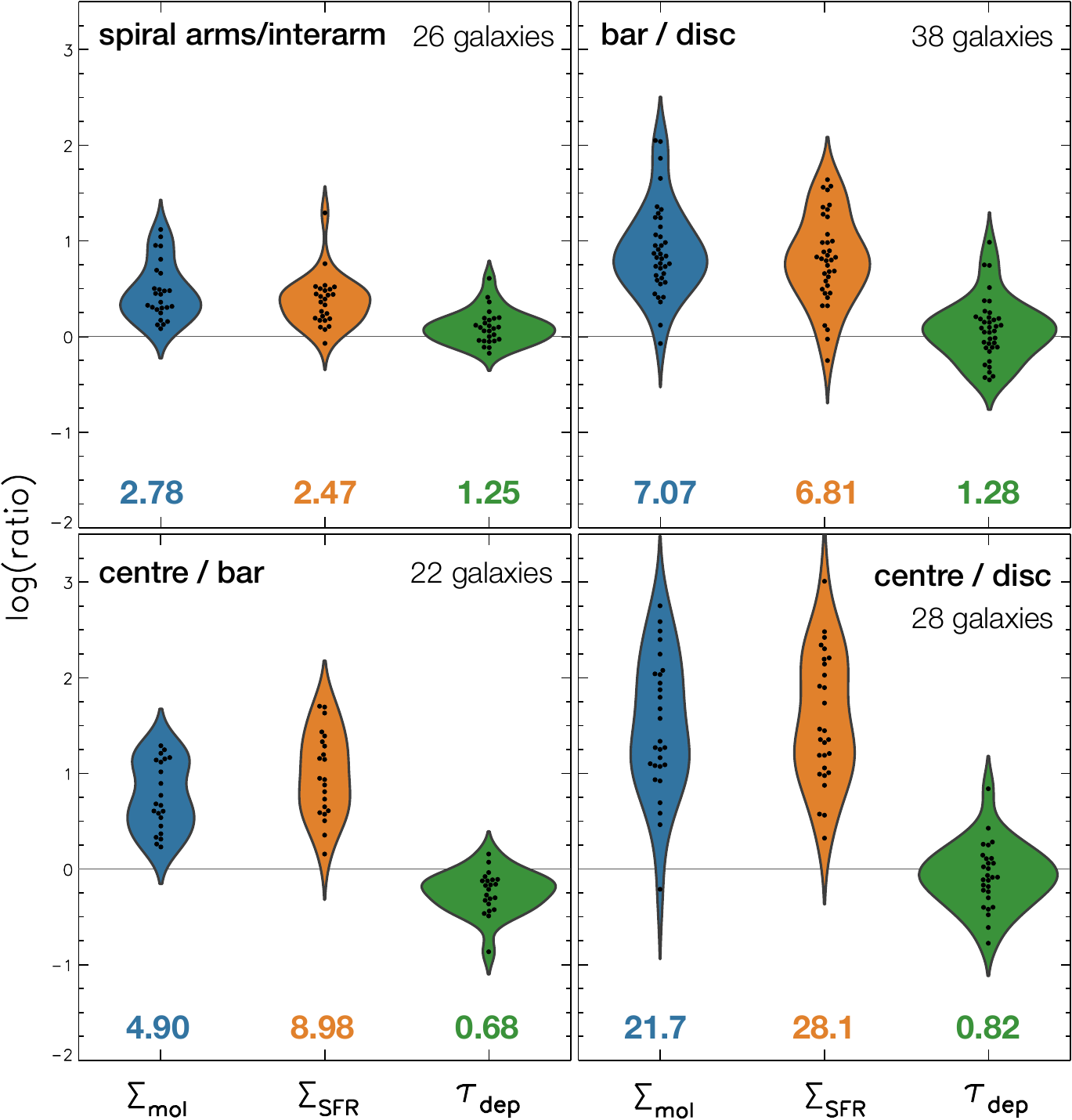}
\end{center}
\caption{Violin plots showing the distribution of contrasts between different pairs of environments in terms of \mod{mean} molecular gas surface density (blue), star formation rate surface density (orange), as well as depletion time (green). These contrasts are calculated within each galaxy (when the pair of structures exists) and each contrast results in a data point, plotted as a black circle. We consider the ratios between spiral arms and interarm (top left); bar and disc (top right), where discs include spiral arms, interarm regions, and discs without spiral masks; centre and bar (bottom left); and centre and disc (bottom right), again including spirals, interarm, and discs without spiral masks. The numbers under the violin plots indicate the median ratio in linear scale. 
\mod{The black dots show the distribution of the ratios; an arbitrary offset is introduced in the horizontal axis to improve visibility.}
}
\label{fig:contrasts}
\end{figure*}

Figure~\ref{fig:contrasts} shows the distribution of surface density contrast within each of the galaxies in pairs of environments. For the arm to interarm contrast, we restricted this ratio to the galactocentric radii spanned by the spiral arms in our masks.
\mod{For each galaxy, we calculated the mean surface density in each environment and defined the contrast as the ratio of those mean surface densities.}
This yields a single number for each contrast per galaxy.
\mod{By examining contrast ratios within galaxies,} we avoid being biased by galaxy-to-galaxy differences driven by large-scale properties; for instance, it is known that galaxies with higher stellar masses tend to have larger global molecular gas and SFR surface densities (e.g.\ \citealt{2007ApJ...670..156D,2008AJ....136.2782L,2011A&A...533A.119E,2016MNRAS.462.1749S}). By focusing on ratios within each galaxy, we are implicitly normalising for galaxy-to-galaxy bulk offsets. For an alternative approach to density contrasts within PHANGS, independent from the environmental masks, we refer the reader to \citet{2021ApJ...913..113M}.

As expected, the arm/interarm ratio of molecular gas and SFR surface densities tends to be greater than~$1$, with a median value of $2.78$ and $2.47$, respectively. These two distributions show a wide range from ${\sim}1$ to ${\sim}10$, and they combine to yield an average arm/interarm ratio of depletion times that can be above or below unity, with a median factor $1.25$. This suggests that in some PHANGS galaxies the spiral arms form stars more efficiently, while in other cases it is the interarm where star formation is on average more efficient. In any case, depletion times in arm and interarm are typically comparable within a given galaxy. We emphasise that our spiral masks are quite wide (typically $1{-}2$\,kpc), in order to accommodate for the small local departures from an ideal log-spiral function (e.g.\ spurs, etc.); a more restrictive mask based on high CO emission along the arm could result in a higher contrast. We also recall that we are performing measurements at kpc-resolution, but the contrast might be resolution-dependent.

Figure~\ref{fig:contrasts} clearly illustrates how centres harbour significantly higher surface densities than bars, and both centres and bars  tend to have substantially higher surface densities than the disc beyond centre and bar (i.e.\ including spiral arms, interarm, and discs without spiral masks). This is expected as a result of the radial trend in surface density, given that molecular gas typically follows a roughly exponential radial profile in disc galaxies \citep[e.g.][]{2001ApJ...561..218R,2008AJ....136.2782L,2011AJ....142...37S}, and star formation closely follows molecular gas. There are a few exceptions where the disc has higher mean surface densities than the bar, but these are cases where ALMA has a limited coverage beyond the bar, so that the interarm measurements mostly come from galactocentric radii similar to the bar.

For centres, the molecular gas and SFR surface density are \mod{between a few times and a few ten times higher than in the bar. This results in depletion times that are most of the times shorter for centres than for bars, with a median ratio of $0.68$}. The shorter depletion times agree with the measurements from \citet{2013AJ....146...19L} in HERACLES, who identify enhanced efficiency in galaxy centres, as well as significant scatter among kpc-sized apertures (see their Fig.\,13); it also agrees with measurements in the Galactic centre \citep{2013MNRAS.429..987L,2014MNRAS.440.3370K}.
When comparing bar to discs, we find ratios of depletion times that are both above and below unity, emphasising galaxy-to-galaxy diversity, even though we are contrasting the same pairs of environments. The median ratios suggest slightly longer depletion times in bars than in discs and slightly shorter depletion times in the centres than in discs, but again with many exceptions.

\section{Discussion} 
\label{Sec:discussion}

Molecular gas and star formation in the PHANGS sample are quite evenly distributed across the five environments that we considered: centres, bars, spiral arms, interarm, and discs without spiral masks. This is in stark contrast with the area covered by these environments, which is tiny for centres ($0.7$\% of total), whereas the combined interarm and discs without spiral masks make up $78$\% of the area across the PHANGS sample. This difference between area and molecular gas or SFR agrees with the expectation that certain stellar structures, such as spiral arms, tend to pile up gas, which also results in higher star formation rates without changing the molecular gas depletion time.

\subsection{Depletion time}

We find a strong correlation between molecular gas and SFR surface densities, with a \mod{global slope of $N=0.97$, and similar slopes among environments, but with a slight offset towards higher SFR surface densities at fixed molecular surface density for centres (median $\tau_{\rm dep}=1.2$\,Gyr), and an offset towards slightly lower SFR surface densities for bars (median $\tau_{\rm dep}=2.1$\,Gyr).}
This is in agreement with \citet{2013AJ....146...19L} and many subsequent studies, who find a tight correlation between molecular gas and SFR surface densities in the HERACLES survey \citep{2009AJ....137.4670L}, well described by a power-law with slope $N=1$. This implies that molecular gas forms stars at a roughly constant efficiency across the discs of nearby galaxies close to the main sequence. \citet{2011MNRAS.415...61S} find a mean molecular gas depletion timescale of ${\sim}1$\,Gyr across the COLD GASS sample (confirmed with a more complete sample in \citealt{2017ApJS..233...22S}), and unveiled a trend with stellar mass, from ${\sim}0.5$\,Gyr for galaxies with stellar mass ${\sim}10^{10}$\,$M_\odot$ to ${\sim}3$\,Gyr for galaxies with masses of a few $10^{11}$\,$M_\odot$. \mod{The median stellar mass in our sample is $2.2 \times 10^{10}$\,$M_\odot$; for that stellar mass, we would expect a depletion time of ${\sim}1$\,Gyr based on \citet{2011MNRAS.415...61S}. Therefore, our median depletion time of $1.4$\,Gyr is slightly longer than this, but in reasonable agreement given the large scatter in the data \citep[see also Fig.\,1 in ][]{2021arXiv210407739L}.}

The shorter median depletion times that we find in centres (particularly in the weighted mean) are likely a genuine effect, since a few galaxies in the sample contribute an outstandingly large share of SFR at the centre (as we commented in Sect.~\ref{Sec:piechart}). In any case, it is worth emphasising that the choice of $\alpha_{\rm CO}$ conversion factor, as well as SFR tracers, can especially affect the measurements for centres (see also \citealt{2021A&A...650A.134P} on the impact of $\alpha_{\rm CO}$ and diffuse ionised gas on the Kennicutt--Schmidt and other scaling relations).
On the other hand, the slightly longer median depletion times in spiral arms than in discs without spiral masks could be attributed to a selection effect: more massive galaxies tend to have better delineated spirals, and more massive galaxies tend to have longer depletion times, too \citep{2011MNRAS.415...61S}. Indeed, PHANGS galaxies with spiral masks have a few times higher stellar masses (median $3.6 \times 10^{10}$\,$M_\odot$) than discs without spiral masks (median $1.0 \times 10^{10}$\,$M_\odot$).

To avoid being biased by differences due to large-scale galaxy properties, such as the one that we just commented on, next we focus on relative differences among environments within individual galaxies. Specifically, we discuss the implications of our findings for spiral arms, bars and centres.

\subsection{Spiral arms}

In optical images, spiral arms stand out due to the presence of bright young stars, and they also map to a local accumulation of gas.
However, the long-standing question remains as to whether spiral arms form stars more efficiently. The question dates back to \citet{1985ApJ...288..438E}. There are theoretical reasons to expect star formation to proceed more efficiently in spiral arms, because the potential well of the arm can create a shock that compresses gas and thus enhances star formation \citep{1969ApJ...158..123R,1975ApJ...196..381R,2004MNRAS.349..909G}. Star formation has also been argued to proceed more efficiently in spiral arms as the result of gravitational instabilities due to reduced shear \citep[e.g.][]{1987ApJ...312..626E, 1993prpl.conf...97E,2002ApJ...570..132K}.
This should be applicable to grand-design spirals, but not as much to flocculent galaxies, and therefore differences could be expected among these two kinds of galaxies. But do spiral arms really trigger enhanced star formation?

First of all, our results confirm that spiral arms play a role in accumulating gas locally and thus star formation. We find that spiral arms have typically $2$~times higher molecular gas and SFR surface densities than the interarm (with a wide range of enhancements from less than a factor of~$2$ up to a factor of ${\sim}10$ in linear scale). This agrees with results from the literature, where typical gas surface densities were found to be a few times higher in spiral arms than interarm regions \citep[e.g.][]{1988Natur.334..402V,1993A&A...274..123G,2003PASJ...55..191N,2009A&A...495..795H,2020ApJ...901L...8S}. Following an alternative approach to define azimuthal density contrasts in PHANGS, \citet{2021ApJ...913..113M} show that there is a correlation between the arm/interarm contrasts in molecular gas and stellar mass, consistent with the expectation from compression driven by large-scale dynamical structures (such as spiral arms and bars).

Despite the locally enhanced molecular and SFR surface densities, the overall contribution of arm and interarm to integrated molecular gas mass and SFR is quite similar (Fig.~\ref{fig:pie_charts}), even though this ultimately depends on the exact definition of the arm masks. According to our results, one cannot claim that star formation in spiral galaxies takes place predominantly in spiral arms or that molecular gas resides mostly in spiral arms. This is something that has been highlighted in a few galaxies before \citep[e.g.][]{2010ApJ...725..534F}; our results confirm and quantify this with a larger sample.

Some observations have found increased star formation efficiency in spiral arms by a factor of ${\sim}2{-}3$ relative to the interarm \citep{1987PhDT........11L,1988Natur.334..402V,1990ApJ...356..135L,1996MNRAS.283..251K}. However, \mod{other} studies have questioned this view: \citet{2010ApJ...725..534F} concluded that the enhancement in star formation efficiency in two grand-design and one flocculent galaxy must be very modest, if any (below $10$\%).
\mod{\citet{2012ApJ...757..155R} find a generally flat star formation efficiency across spiral arms and the interarm in NGC\,6946, even though a few specific spiral arm regions appear to have higher efficiency than other arm or interarm structures; a similar conclusion was reached by \citet{2003AJ....126.2831H} for M51 (see also \citealt{2019A&A...625A..19Q}).}
\citet{2016ApJ...827..103K} also confirmed with higher-resolution data that there are no significant differences in the molecular gas depletion times between arm and interarm in NGC\,628.
\mod{Using COLD GASS and HERACLES data, \citet{2014MNRAS.443.1329H} show that $\tau_{\rm dep}$ has a primary dependence on specific SFR (both on integrated galaxy scales and for kpc-sized disc apertures). \citet{2015MNRAS.450.1375H} further show that, at fixed specific SFR, $\tau_{\rm dep}$ tends to be relatively longer for galaxies with spiral arms. Yet, spiral arms span a wide range of molecular gas depletion times, with a large overlap among different stellar structures. These findings from \citet{2015MNRAS.450.1375H} broadly agree with our results and highlight the diverse range of conditions for similar morphological features across nearby galaxies. This could partially stem from the fact that we observe different galaxies at different evolutionary stages, in addition to intrinsic structural differences.}

Milky Way studies have also reached similar conclusions: \citet{2013MNRAS.431.1587E}, \citet{2018MNRAS.479.2361R} and \citet{2020ApJ...904..147U} do not find evidence for enhanced star formation efficiency in Galactic spiral arms. More specifically, observations do not support a strong connection between shear and star formation efficiency in the Milky Way and M51 \citep{2012ApJ...758..125D,2013ApJ...779...45M}; therefore, we should not necessarily expect increased efficiencies in spiral arms.

In this context, our results confirm that, globally, spiral arms do not result in shorter depletion times (Fig.~\ref{fig:contrasts}). However, this does not mean that depletion times are always identical in the arm and interarm; as a matter of fact, in some cases the depletion times are longer in spiral arms and in some other cases they are shorter. These fluctuations in arm/interarm depletion times are typically within a factor of~$2$ above or below unity, with a median centred at $1.25$. It is possible to find individual cases of shorter depletion times in spiral arms, in agreement with some studies from the literature, but there are more cases where the efficiency is actually lower in spiral arms. So far, these kinds of measurements have been considered in individual case studies, but here we confirm with a much larger sample that spiral arms on kpc scales do not typically enhance the star formation efficiency relative to the interarm. Our findings confirm the idea that spiral arms act to accumulate gas and star formation, but do not on average trigger more efficient star formation. 

Simulations essentially agree with what we find observationally. \citet{2006MNRAS.365...37B}, \citet{2009MNRAS.396.1579D}, and \citet{2011MNRAS.417.1318D} used hydrodynamical simulations of a galaxy to show that, while molecular gas tends to accumulate in spiral arms, producing more massive GMCs, these do not result in an enhanced star formation efficiency. The higher velocity dispersion of these structures makes them less susceptible to collapse, and far from triggering star formation, spiral arms simply play the role of piling up the gas (and therefore star formation), without increasing the efficiency at which the gas forms new stars \citep[see also e.g.][]{2020ApJ...898...35K,2020MNRAS.492.2973T}. \mod{\citet{2017ApJ...845..133S} also studied how gas cools but simultaneously increases its velocity dispersion as it enters a spiral arm, and these two counter-balancing effects can result in a similar depletion time between arm and interarm.}

One can also wonder if flocculent discs behave differently from stronger (typically grand-design) spirals. \citet{2008AJ....136.2846B} and \citet{2008AJ....136.2782L} did not find important differences in the star formation efficiency of strongly armed spirals as opposed to weakly armed spirals. Similarly, \citet{2010ApJ...725..534F} did not find significant differences between two grand-design and one flocculent spiral that they examined; the simulations of \citet{2020MNRAS.492.2973T} also point to comparable star formation efficiency in a grand-design and a flocculent spiral disc. In our study, Fig.~\ref{fig:histo_surfdens} could convey the impression that galaxies with spiral masks (predominantly grand-design) and the rest of discs (closer to flocculent) behave \mod{slightly} differently but, as we discussed above, this is likely a trend driven by stellar mass. This emphasises that one has to be careful when contrasting environments from different galaxies.

Our spiral masks are relatively broad by construction (with a typical width of $1{-}2$\,kpc), as our goal is that they encompass the different tracers that we are interested in (NIR, CO, H$\alpha$, etc.), and accommodate local irregularities relative to a log-spiral curve, such as spurs. Therefore, a more restrictive definition of the spiral arm width, following the high surface density ridge of molecular gas, would naturally result in a higher arm/interarm ratio for the gas surface density (and possibly star formation). However, when it comes to depletion times, if they are intrinsically shorter in the spiral arm than interarm, we do not expect a spiral mask that is slightly too broad to invert this trend: the presence of some interarm within the spiral mask could slightly dilute the shorter depletion times in the arm, but on average they should still remain shorter than in the interarm. Since we observe the opposite, with most of our galaxies displaying longer depletion times in the arm, we conclude that our conservatively wide spiral masks should not affect this conclusion. It is also worth mentioning that offsets among different tracers are expected in spiral arms \citep[e.g.][]{2013ApJ...779...42S,2013MNRAS.433.1837V,2016ApJ...827..103K,2017ApJ...845...78C,2017MNRAS.465..460E}, and this can compromise the measured star formation efficiency locally for a given position. If arms trigger a time sequence, one can expect regions of high apparent efficiency where star formation concentrates. Our measurements intentionally average over kpc-sized hexagonal apertures in order to minimise such local biases (azimuthal offsets are typically smaller than 1\,kpc), but this information can be useful in other contexts. 

\subsection{Bars}

A number of observations support the idea that star formation is suppressed in strong bars \citep[e.g.][]{2009A&A...501..207J,2010ApJ...721..383M,2014PASJ...66...46H,2016MNRAS.456.2848H}. Numerical simulations have also found suppression of star formation along bars \citep[e.g.][]{1982ApJ...255..458T,1992MNRAS.259..345A,2015MNRAS.446.2468E,2017MNRAS.465.3729S}, suggesting as possible mechanisms for this effect a combination of the bar quickly funnelling gas towards the centre  (removing the fuel for star formation) and strong shear making the gas less prone to form stars.

PHANGS provides a homogeneous census of molecular gas and star formation across a representative sample of local barred galaxies.
Our results suggest that bars are not quiescent structures devoid of molecular gas; in fact, we find that molecular gas and SFR surface densities in bars are typically higher than in the disc (or interarm) and comparable to spiral arms. Depletion times in bars are also comparable to spiral arms or interarm. The distribution of surface densities is broad, with a range of values that suggests large diversity within individual bars and also among galaxies.
\mod{Using COLD GASS and HERACLES observations, \citet{2015MNRAS.450.1375H} demonstrate that barred galaxies tend to have shorter molecular gas depletion times at fixed specific SFR than galaxies without bars, but barred regions show a wide range of values spanning more than a factor of 10. Indeed, in our observations, one can also}
find apertures within bars where star formation is suppressed, but also other locations where star formation is enhanced. Indeed, it is known that bar ends can host significant and efficient star formation, and bar ends are included within our bar masks \citep[e.g.][]{1991ApJ...381..118K, 1997A&A...326..449M,2017A&A...597A..85B}. Converging flows and cloud-cloud collisions have been suggested as possible triggers of star formation in bar ends \citep{1976ApJ...209..466L,2000ApJ...536..173T,2009ApJ...696L.115F,2009ApJ...700..358T,2013ApJ...774L..31I,2015MNRAS.454.3299R} and even along the bar \citep{2020MNRAS.497.5024S,2020MNRAS.499.4455T}. The star formation distribution in bars seems to depend on host galaxy morphology, with late-type, gas-rich galaxies showing more extended star formation along the bar, possibly as a result of decreased shear \citep{2020A&A...644A..38D}.

While Fig.~\ref{fig:histo_surfdens} plots together sight lines from different galaxies, Fig.~\ref{fig:contrasts} confirms that the high surface density in bars relative to the disc or interarm environment is not a false impression due to mixing sight lines across galaxies; when we compute the ratio of surface densities within individual galaxies we also find that bars nearly always harbour higher surface densities than the disc beyond the bar.
These high surface density bars represent an important contribution to the high overall gas surface densities seen in the inner parts compared to the outer parts of galaxies, and may largely reflect the overall exponential appearance of gas discs \citep[e.g.][]{2008AJ....136.2782L}.
It is also worth emphasising that the magnitude of the bar/disc or centre/disc contrast depends on the ALMA field of view: if the ALMA coverage extends further out, the average surface density of the disc will drop, making the corresponding ratio higher. In any case, the depletion time circumvents this issue, as it considers the ratio of molecular gas and SFR surface densities, implicitly normalising for any common basic radial trend in surface densities. Thus, the comparable depletion times in bars and discs or interarm confirms that, on average, molecular gas is not \mod{systematically} forming stars less efficiently in the bar environment. This result may have been unexpected, given the widespread literature claiming that bars are star formation deserts; our results suggest that bars are complex sites where processes resulting both in relatively strong and feeble star formation are at play.

For each barred galaxy, we computed the CO-weighted average galactocentric radius inside the bar footprint (excluding centres), normalised to the bar length, and found values typically ranging $R_\textrm{CO-weighted}/R_{\rm bar} \sim 0.4{-}0.7$, with a few outliers around $0.9$ or higher.
This means that most of the cold gas in our sample is not sitting at the bar ends, and rather distributed across the entire bar. Indeed, the expectation for relatively enhanced or suppressed star formation in bars will likely depend on where the bulk of molecular gas is located (high SFE in bar ends, low SFE along dust lanes). This measurement of the CO-weighted radius tells us that the PHANGS sample contains galaxies with gas located throughout the bar, and this helps explain the diversity in observed SFEs. 
\mod{As shown by \citet{2019A&A...627A..26N}, galaxies of a given Hubble type can host both star-forming and non-star-forming bars.}
Globally, these findings suggest that, rather than static stellar structures that quickly funnel gas towards centres, bars typically keep growing in stellar mass as the molecular gas that they harbour forms new stars. This process sustained in time agrees with the findings for relatively extended star formation histories in bars and barlenses \citep[e.g.][]{2018A&A...618A..34L,2019MNRAS.482..506G,2019A&A...627A..26N,2020A&A...637A..56N}. In any case, our results call for more detailed studies of bars within PHANGS.

\subsection{Centres}

Early studies already reported higher molecular gas masses and associated gas mass surface densities in the central kpc of barred galaxies compared to those of unbarred ones \citep{1999ApJ...525..691S,2005ApJ...630..837J,2005ApJ...632..217S,2007PASJ...59..117K}. A similar enhancement has also been observed in our Milky Way centre \citep{2001ApJ...562..348O,2016MNRAS.457.2675H}.
Recently, \citet{2018ApJ...860..172S,2020ApJ...901L...8S} showed that the centres of barred galaxies in PHANGS harbour exceptionally high molecular gas surface density and velocity dispersion at ${\sim}100$\,pc scales. The former is attributable to the large-scale, bar-driven inflow; the latter is hypothesised to stem from inflow or unresolved orbital motions as well as the stronger local star formation feedback.
These higher gas masses or surface densities are often accompanied by higher SFR surface densities as traced by high resolution H$\alpha$ imaging and/or imaging spectroscopy \citep{2005ApJ...630..837J,2019MNRAS.484.5192C}. This difference in SFR between galaxy centres and galactic discs is also consistent with the findings from radio continuum imaging by the Star Formation in Radio Survey (SFRS: \citealt{2012ApJ...761...97M}) which is insensitive to attenuation effects, and which suggest that circumnuclear star formation is characterised by a more extended or continuous star formation history compared to the younger and more transient star formation occurring in galaxy discs \citep{2011ApJ...737...67M,2018ApJS..234...24M,2020ApJS..248...25L}.
Our measurements confirm remarkably high surface densities in centres at kpc scales, up to $\Sigma_{\rm mol} \sim 1000$\,$M_\odot$\,pc$^{-2}$ and $\Sigma_{\rm SFR} \sim 1$\,$M_\odot$\,yr$^{-1}$\,kpc$^{-2}$. Compared to other environments, the typical molecular gas and SFR surface densities in centres are between $10$ and $100$ times higher. However, it is possible to find centres with surface densities lower than those of spirals or discs. This highlights the importance of morphology or galactic environment (e.g. bulges, bars) in gas organisation and other global host galaxy properties such as stellar mass \citep[e.g.][]{2017ApJ...846..159B,2020ApJ...901L...8S}.

If we focus on depletion times, we find that not all centres are equally efficient at forming stars. The distribution is very broad and strongly overlaps with other environments, implying that central depletion times can be a few times shorter or longer than in other environments. This highlights again the wide range of conditions and star formation efficiency in centres and agrees with previous studies \citep[e.g.][]{2013AJ....146...19L,2018ApJ...861L..18U}. The median depletion time that we find across PHANGS centres (1.2\,Gyr) is slightly longer than measurements in the Galactic centre (${\sim}0.6$\,Gyr; \citealt{2013MNRAS.429..987L,2014MNRAS.440.3370K}).
We caution again regarding the comparison of different galaxies, which is why a ratio of depletion times in centres to discs might be more meaningful; this ratio is often (but now always) below unity, with a median $0.82$.
Based on observations from HERACLES, \citet{2013AJ....146...19L} find a median central-to-disc $\tau_{\rm dep}$ ratio of 0.6 with 0.35\,dex scatter for a spatially varying $\alpha_{\rm CO}$ conversion factor (similar to our fiducial choice of $\alpha_{\rm CO}$ here), which goes in line with our results.

\begin{figure*}[t]
\begin{center}
\includegraphics[trim=0 0 0 0, clip,width=0.9\textwidth]{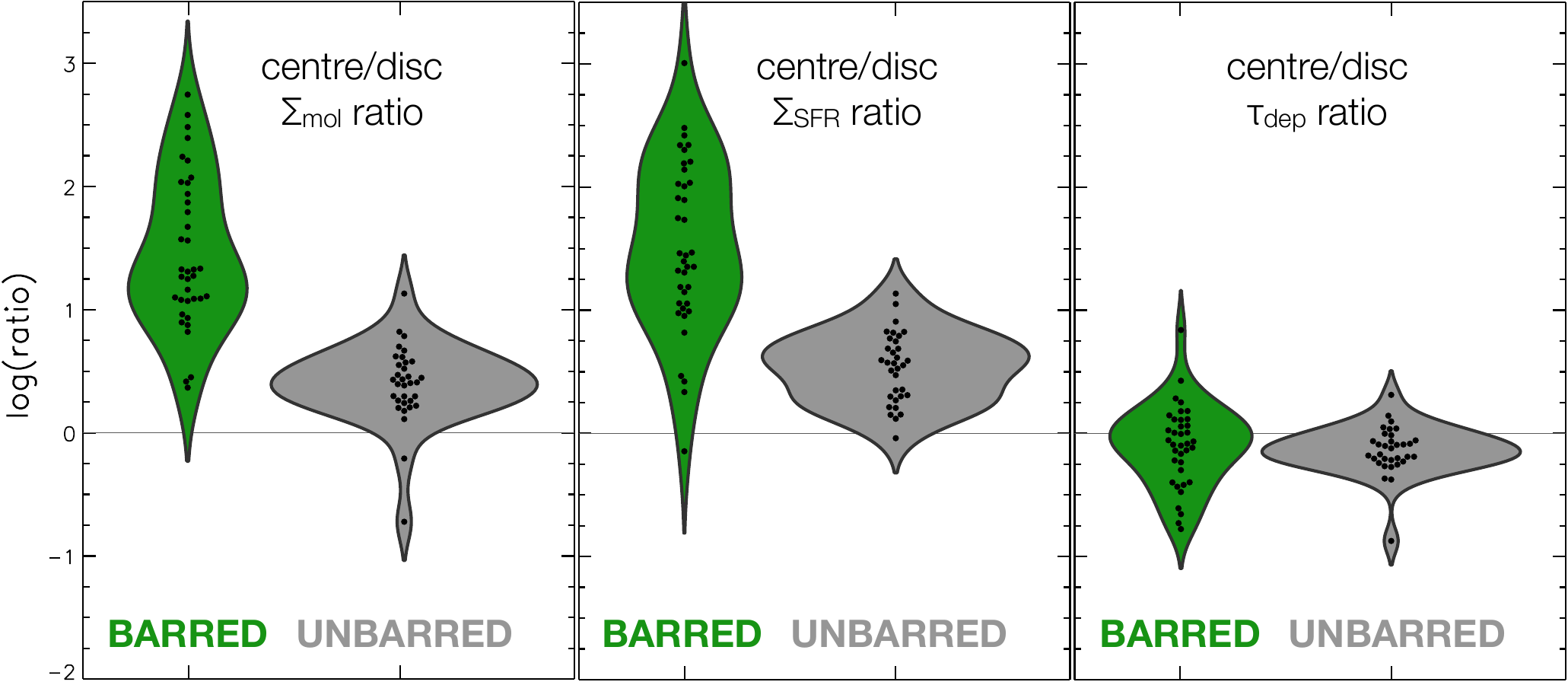}
\end{center}
\caption{Violin plots showing the contrast in \mod{molecular gas and star formation rate surface density, as well as} depletion time, between centres and discs. \mod{The sample is} split into barred (green) and unbarred galaxies (grey).
}
\label{fig:ratio_cen-disc}
\end{figure*}

In Figure~\ref{fig:ratio_cen-disc} we consider whether centres of barred galaxies have intrinsically different \mod{properties} from centres of unbarred galaxies, relative to the outer discs. The most striking feature is that barred galaxies show \mod{more elevated surface densities, with a markedly wider distribution (the standard deviations in linear scale are $122$ versus $2.4$ and $176$ versus $2.9$, for $\Sigma_{\rm mol}$ and $\Sigma_{\rm SFR}$ in barred and unbarred galaxies, respectively)}. Despite this remarkable difference, the median ratios of depletion times are very similar ($0.82$ and $0.75$ for barred and unbarred galaxies, respectively); \mod{the distribution of depletion times is slightly wider in barred galaxies (standard deviation of $1.12$ versus $0.34$), but the difference is nowhere as extreme as for surface densities.}
\mod{Here, we consider a central measurement for all galaxies, either the entire `centre' mask, or the kpc-sized hexagonal aperture at $R=0$ for all the galaxies without an explicit `centre' mask. We do this} 
in order to avoid being biased towards galaxies with central (often bar-driven) structures. 
As the outer reference for the ratio, we consider either discs without spiral masks or spirals plus interarm.

The large difference in the scatter of \mod{$\Sigma_{\rm mol}$ and $\Sigma_{\rm SFR}$ (and, to a lesser extent, $\tau_{\rm dep}$)} is in line with the findings from \citet{2007PASJ...59..117K} and \citet{2020ApJ...901L...8S}, who show that centres of barred galaxies behave very differently from the centres of unbarred galaxies in terms of surface densities and velocity dispersion. The higher dispersion in \mod{centres of} barred galaxies is expected on theoretical grounds, as bars tend to funnel gas episodically to the centre, and it cycles through bursts of star formation and more quiescent phases \citep[e.g.][]{2013ApJ...769..100S,2014MNRAS.440.3370K,2017MNRAS.466.1213K,2020MNRAS.497.5024S}. This could explain the diversity that we observe. By comparison, the centres of unbarred galaxies are presumably more uniform and stable as a function of time, as there is generally not such a strong driver of (intermittent) gas inflow towards the centre \citep[e.g.][]{2005A&A...441.1011G,2009ApJ...692.1623H}.
Thus, on average, the bar-driven inflows do not seem to enhance the star formation efficiency permanently, but rather trigger intermittent fluctuations (both up and down) in the depletion time. These fluctuations could be driven indirectly by bar-built structures such as dust lanes and nuclear rings, as they evolve with time.

Centres are dynamically complex regions characterised by high shear. Yet, for a given star formation threshold, we can expect the gas that reaches higher densities to be more strongly self-gravitating and thus presumably more efficiently at forming stars \citep{2018ApJ...854..100M,2020ApJ...892...73M,2019MNRAS.484.5734K}. Specifically, gas in a compact nuclear ring is more likely to be forming stars than gas that is more uniformly distributed, at lower densities. Our kpc-scale measurements average out many of these small-scale effects, but we plan to revisit this topic in detail in future PHANGS publications. Indeed, centres are a mixed bag with potentially very different sub-structures that can only be resolved at high resolution.

\section{Summary and conclusions} 
\label{Sec:concl}

We present a catalogue of two-dimensional masks for PHANGS galaxies delimiting different morphological environments. The masks are based on {\it Spitzer} $3.6$\,$\mu$m emission, and they identify stellar structures such as discs, bulges, bars, spiral arms, rings and lenses (Fig.~\ref{fig:intro} shows some examples). We mostly rely on archival {\it Spitzer} imaging (largely from the S$^4$G survey), but we also carried out new {\it Spitzer} observations of four PHANGS galaxies that did not have archival IRAC data (NGC\,2283, NGC\,2835, NGC\,3059 and NGC\,3137), following an observing and data reduction strategy analogous to S$^4$G.

Bulges and discs are defined based on photometric decompositions of near-infrared images via GALFIT or similar tools (\citealt{2015ApJS..219....4S} for S$^4$G galaxies; \citealt{2004MNRAS.355.1251L} or new fits otherwise). We define the outer edge of bulges and discs as twice their effective radius. The sizes of bars, rings and lenses are defined visually and their ellipticity is measured via ellipse fitting (\citealt{2015A&A...582A..86H} for S$^4$G; other measurements from the literature otherwise). We identify spiral arms as peaks on unsharp-masked $3.6$\,$\mu$m images followed by log-spiral fits in polar coordinates (\citealt{2015A&A...582A..86H} and new measurements), with a width that is empirically established based on the spatial distribution of CO emission.

Even though environments often overlap (e.g.\ a nuclear ring often lies within the bar footprint), we propose a simple strategy to uniquely assign pixels to a dominant environment (Fig.~\ref{fig:non-overlapping_masks}). Like this, the environmental masks can be simplified to `centre', `bar', `spiral arms', and `interarm' (or a generic `disc' for galaxies without spiral masks), and this is the approach that we follow in this paper.
Our main results are the following:

\begin{enumerate}

\item As much as three quarters of the PHANGS--ALMA (deprojected) area corresponds to interarm and discs without spirals. However, molecular gas and star formation are quite evenly distributed among centres, bars, spiral arms, interarm, and discs without spirals (Fig.~\ref{fig:pie_charts}). This highlights the relevance of centres, despite their very limited area ($0.7$\% of total), and that star formation is not taking place predominantly in spiral arms, since bars and interarm have similar integrated molecular masses and SFRs.

\item There is a trend for components at larger radii to have lower surface densities, as expected (Fig.~\ref{fig:SurfDens_vs_Rnorm}). We find a large range of surface density at fixed normalised radius (up to $2{-}3$~dex) even within a given environment. On top of this, environments largely overlap in the $\Sigma {-} R/R_{25}$ planes, with an offset of spiral arms towards higher surface densities (both $\Sigma_{\rm mol}$ and $\Sigma_{\rm SFR}$) \mod{relative to the interarm} at fixed normalised radius.

\item Our measurements follow a strong correlation between star formation rate and molecular gas surface densities (molecular Kennicutt--Schmidt relation). The overall slope is very close to unity ($N = 0.97 \pm 0.06$), and the slopes fitted independently for each environment are also compatible with unity within the uncertainties. The largest vertical offset with respect to the global fit is found in centres ($+0.23$\,dex), pointing to slightly shorter average depletion times.

\item Centres harbour remarkably high surface densities (median $\Sigma_{\rm mol} = 106 $\,$M_\odot$\,pc$^{-2}$ and $\Sigma_{\rm SFR} = 0.07$\,$M_\odot$\,yr$^{-1}$\,kpc$^{-2}$), and a wide range of depletion times (with a median of $1.18$\,Gyr). \mod{Surface densities in centres of barred galaxies are more elevated (and show a much larger dispersion) relative to the disc, but the centre/disc ratio of depletion times is similar in barred and unbarred galaxies.} This could be due to bars fuelling gas episodically to the centre, which cycles through bursts of star formation and more quiescent phases. Thus, on average, the bar-driven inflows do not seem to enhance the star formation efficiency permanently, but rather induce intermittent fluctuations.

\item We find high molecular gas and star formation rate surface densities in bars, comparable to spiral arms and higher than interarm or discs without spirals (median $\Sigma_{\rm mol} = 11.5$\,$M_\odot$\,pc$^{-2}$ and $\Sigma_{\rm SFR} = 0.006$\,$M_\odot$\,yr$^{-1}$\,kpc$^{-2}$). Contrary to claims of suppressed star formation in bars, we do not find evidence for \mod{systematically} longer depletion times in bars, and our measurements rather support the idea that there is large diversity in the star-forming conditions in bars within and among galaxies (probably mediated by bar-built structures such as dust lanes and nuclear rings).

\item When averaged inside our masks, molecular gas and star formation surface densities tend to be higher in spiral arms than in the interarm (covering a wide range of ratios from ${\sim}1$ to ${\sim}10$, with medians $2.8$ and $2.5$, respectively). Contrary to some previous claims, and in agreement with other studies from the literature, we do not find evidence for systematically enhanced star formation efficiencies in spiral arms. There are individual galaxies where star formation is more efficient in arms, but also many others where star formation is less efficient. Therefore, spiral arms seem to act to pile up gas, and consequently star formation, but do not preferentially lead to shorter depletion times.

\end{enumerate} 

In conclusion, our observations suggest that there can be substantial diversity in molecular gas and star formation within and among galaxies, linked to galactic structure. We plan to revisit some of these topics in detail in future publications, exploiting the rich multi-wavelength data sets produced by PHANGS. In this context, the environmental masks that we have presented here should help illuminate some of the physical processes by which stellar structures regulate star formation in the local Universe.

\small  
%
\begin{acknowledgements}   
This work was carried out as part of the PHANGS collaboration. \mod{We would like to thank the anonymous referee for very detailed and valuable comments that helped us improve the manuscript.}
We would also like to thank Seppo Laine \mod{and the whole \textit{Spitzer} Helpdesk} for advice regarding IRAC data reduction.

This work is based on observations and archival data obtained with the \textit{Spitzer} Space Telescope, which is operated by the Jet Propulsion Laboratory, California Institute of Technology under a contract with NASA.
This paper makes use of the following ALMA data:  ADS/JAO.ALMA\#2012.1.00650.S,  
ADS/JAO.ALMA\#2013.1.00803.S,  
ADS/JAO.ALMA\#2013.1.01161.S,  
ADS/JAO.ALMA\#2015.1.00121.S,  
ADS/JAO.ALMA\#2015.1.00782.S,  
ADS/JAO.ALMA\#2015.1.00925.S,  
ADS/JAO.ALMA\#2015.1.00956.S,  
ADS/JAO.ALMA\#2016.1.00386.S, 
ADS/JAO.ALMA\#2017.1.00886.L,  
ADS/JAO.ALMA\#2018.1.01651.S.  
ALMA is a partnership of ESO (representing its member states), NSF (USA) and NINS (Japan), together with NRC (Canada), MOST and ASIAA (Taiwan), and KASI (Republic of Korea), in cooperation with the Republic of Chile. The Joint ALMA Observatory is operated by ESO, AUI/NRAO and NAOJ. The National Radio Astronomy Observatory is a facility of the National Science Foundation operated under cooperative agreement by Associated Universities, Inc.

MQ acknowledges support from the research project PID2019-106027GA-C44 from the Spanish Ministerio de Ciencia e Innovaci\'on.
ES, DL, HAP, TS, and TGW acknowledge funding from the European Research Council (ERC) under the European Union’s Horizon 2020 research and innovation programme (grant agreement No. 694343).
The work of JS and AKL is partially supported by the National Science Foundation (NSF) under Grants No.\ 1615105, 1615109, and 1653300, as well as by the National Aeronautics and Space Administration (NASA) under ADAP grants NNX16AF48G and NNX17AF39G.
RSK and SCOG acknowledge financial support from the German Research Foundation (DFG) via the collaborative research centre (SFB 881, Project-ID 138713538) “The Milky Way System” (subprojects A1, B1, B2, and B8). They also acknowledge funding from the Heidelberg Cluster of Excellence ``STRUCTURES'' in the framework of Germany’s Excellence Strategy (grant EXC-2181/1, Project-ID 390900948) and from the European Research Council via the ERC Synergy Grant ``ECOGAL'' (grant 855130). 
HS and EL acknowledge financial support from the European
Union's Horizon 2020 research and innovation programme under the Marie
Sklodowska-Curie grant agreement No.\ 721463 to the SUNDIAL ITN network, and by the Academy of Finland grant No.\ 297738.
IB acknowledges funding from the European Research Council (ERC) under the European Union’s Horizon 2020 research and innovation programme (grant agreement No.726384/Empire).
MC and JMDK gratefully acknowledge funding from the Deutsche Forschungsgemeinschaft (DFG) in the form of an Emmy Noether Research Group (grant number KR4801/1-1) and the DFG Sachbeihilfe (grant number KR4801/2-1), as well as from the European Research Council (ERC) under the European Union’s Horizon 2020 research and innovation programme via the ERC Starting Grant MUSTANG (grant agreement number 714907).
CE acknowledges funding from the Deutsche Forschungsgemeinschaft (DFG) Sachbeihilfe, grant number BI1546/3-1.
CMF is supported by the National Science Foundation under Award No. 1903946 and acknowledges funding from the European Research Council (ERC) under the European Union’s Horizon 2020 research and innovation programme (grant agreement No. 694343).
AGR acknowledges support from the Spanish funding grants AYA2016-79006-P (MINECO/FEDER) and PID2019-108765GB-I00 (MICINN). 
AH was supported by the Programme National Cosmology et Galaxies (PNCG) of CNRS/INSU with INP and IN2P3, co-funded by CEA and CNES, and by the Programme National “Physique et Chimie du Milieu Interstellaire” (PCMI) of CNRS/INSU with INC/INP co-funded by CEA and CNES.
CH and JP acknowledge support from the Programme National “Physique et Chimie du Milieu Interstellaire” (PCMI) of CNRS/INSU with INC/INP co-funded by CEA and CNES.
KK gratefully acknowledges funding from the German Research Foundation (DFG) in the form of an Emmy Noether Research Group (grant number KR4598/2-1, PI Kreckel).
ER acknowledges the support of the Natural Sciences and Engineering Research Council of Canada (NSERC), funding reference number RGPIN-2017-03987.
AU acknowledges support from the Spanish funding grants AYA2016-79006-P (MINECO/FEDER), PGC2018-094671-B-I00 (MCIU/AEI/FEDER), and PID2019-108765GB-I00 (MICINN). 
EJW is funded by the Deutsche Forschungsgemeinschaft (DFG, German Research Foundation) -- Project-ID 138713538 -- SFB 881 (``The Milky Way System'', subproject P2).
\end{acknowledgements}

\newpage

\normalsize
\appendix
\section{Sanity checks}
\label{sec:Appendix1}

\begin{table*}[t!]
\small
\begin{center}
\caption[h!]{Sanity checks showing how results are affected by different choices of some specific parameters.}
\begin{tabular}{llcccccc}
\hline\hline
 & & centre & bar & spiral & interarm & disc & all  \\
   \hline
   \hline
\multicolumn{2}{l}{Area [kpc$^2$]} & 92 (0.66\,\%) & 1467 (10.5\,\%) & 1495 (10.7\,\%) & 4620 (33.0\,\%) & 6296 (45.0\,\%) & 13972 (100\,\%) \\
\multicolumn{2}{l}{H$_2$ mass [$10^{10}$\,$M_\odot$]} & 2.32 (17.4\,\%) & 2.64 (19.8\,\%) & 2.45 (18.4\,\%) & 2.25 (16.8\,\%) & 3.66 (27.4\,\%) & 13.3 (100\,\%) \\
\multicolumn{2}{l}{SFR [$M_\odot$\,yr$^{-1}$]} & 23.9 (25.2\,\%) & 15.6 (16.4\,\%) & 15.6 (16.5\,\%) & 13.1 (13.9\,\%) & 26.2 (27.7\,\%) & 94.6 (100\,\%) \\
   \hline
\multicolumn{2}{l}{Area [kpc$^2$], $\rm FoV > R_{25}$} & 66 (0.59\,\%) & 1215 (10.8\,\%) & 1094 (9.77\,\%) & 3352 (29.9\,\%) & 5468 (48.8\,\%) & 11197 (100\,\%) \\
\multicolumn{2}{l}{H$_2$ mass [$10^{10}$\,$M_\odot$], $\rm FoV > R_{25}$} & 1.17 (12.8\,\%) & 1.81 (19.8\,\%) & 1.52 (16.6\,\%) & 1.46 (15.9\,\%) & 3.17 (34.6\,\%) & 9.16 (100\,\%) \\
\multicolumn{2}{l}{SFR [$M_\odot$\,yr$^{-1}$], $\rm FoV > R_{25}$} & 14.3 (21.4\,\%) & 11.2 (16.8\,\%) & 8.68 (12.9\,\%) & 8.01 (11.9\,\%) & 24.5 (36.7\,\%) & 66.8 (100\,\%) \\
\hline
\hline
   \noalign{\smallskip}
\multirow{3}{*}{$\Sigma_\mathrm{mol}/(M_\odot~\mathrm{pc^{-2}})$ } &  median & $106.1_{-82.97}^{+225.5}$ & $11.51_{-8.620}^{+27.45}$ & $9.903_{-6.637}^{+16.91}$ & $4.492_{-2.924}^{+6.079}$ & $4.120_{-2.631}^{+7.610}$ & $5.001_{-3.290}^{+10.57}$ \\
 &  mean & $159.4$ & $20.92$ & $14.97$ & $6.589$ & $6.671$ & $10.03$ \\
 &  weighted & $340.8$ & $56.14$ & $29.82$ & $14.66$ & $15.15$ & $59.26$ \\
   \noalign{\smallskip}
\multirow{3}{*}{$\Sigma_{\rm SFR}/(M_\odot~\mathrm{yr^{-1}}~\mathrm{kpc^{-2}})$ } &  median & $0.0739_{-0.0592}^{+0.3406}$ & $0.0057_{-0.0045}^{+0.0171}$ & $0.0055_{-0.0037}^{+0.0100}$ & $0.0026_{-0.0016}^{+0.0034}$ & $0.0025_{-0.0015}^{+0.0052}$ & $0.0029_{-0.0018}^{+0.0061}$ \\
 &  mean & $0.2079$ & $0.0156$ & $0.0094$ & $0.0042$ & $0.0046$ & $0.0074$ \\
 &  weighted & $0.4956$ & $0.0596$ & $0.0198$ & $0.0094$ & $0.0108$ & $0.0715$ \\
   \noalign{\smallskip}
\multirow{3}{*}{$\tau_{\rm dep}/\mathrm{(Gyr)}$} &  median & $1.177_{-0.652}^{+0.650}$ & $2.102_{-1.041}^{+1.637}$ & $1.788_{-0.735}^{+1.038}$ & $1.677_{-0.872}^{+1.369}$ & $1.565_{-0.746}^{+1.245}$ & $1.671_{-0.824}^{+1.309}$ \\
 &  mean & $1.188$ & $2.432$ & $2.007$ & $2.459$ & $2.237$ & $2.296$ \\
 &  weighted & $0.978$ & $2.238$ & $1.949$ & $2.528$ & $2.106$ & $2.083$ \\
\hline
   \noalign{\smallskip}
\multirow{3}{*}{$\Sigma_\mathrm{mol}$   thr $70$\%} &  median & $105.3_{-83.50}^{+111.7}$ & $11.45_{-8.599}^{+27.31}$ & $10.30_{-6.958}^{+15.99}$ & $4.588_{-3.000}^{+6.246}$ & $4.109_{-2.625}^{+7.627}$ & $5.119_{-3.376}^{+11.02}$ \\
 &  mean & $142.7$ & $20.94$ & $14.92$ & $6.829$ & $6.688$ & $10.32$ \\
 &  weighted & $314.9$ & $57.02$ & $29.06$ & $16.04$ & $15.25$ & $58.31$ \\
   \noalign{\smallskip}
\multirow{3}{*}{$\Sigma_{\rm SFR}$   thr $70$\%} &  median & $0.0732_{-0.0601}^{+0.3272}$ & $0.0058_{-0.0047}^{+0.0168}$ & $0.0056_{-0.0037}^{+0.0100}$ & $0.0027_{-0.0017}^{+0.0035}$ & $0.0025_{-0.0015}^{+0.0053}$ & $0.0030_{-0.0019}^{+0.0064}$ \\
 &  mean & $0.1913$ & $0.0161$ & $0.0094$ & $0.0043$ & $0.0047$ & $0.0078$ \\
 &  weighted & $0.4603$ & $0.0634$ & $0.0191$ & $0.0101$ & $0.0108$ & $0.0705$ \\
   \noalign{\smallskip}
\multirow{3}{*}{$\tau_{\rm dep}$   thr $70$\%} &  median & $1.174_{-0.694}^{+0.653}$ & $2.077_{-1.045}^{+1.616}$ & $1.772_{-0.732}^{+0.981}$ & $1.685_{-0.876}^{+1.332}$ & $1.564_{-0.744}^{+1.245}$ & $1.671_{-0.820}^{+1.289}$ \\
 &  mean & $1.171$ & $2.391$ & $1.973$ & $2.433$ & $2.229$ & $2.271$ \\
 &  weighted & $0.986$ & $2.188$ & $1.944$ & $2.483$ & $2.100$ & $2.055$ \\
\hline
    \noalign{\smallskip}
\multirow{3}{*}{$\Sigma_\mathrm{mol}$ thr $90$\%} &  median & $118.9_{-93.65}^{+267.4}$ & $11.31_{-8.462}^{+27.26}$ & $10.29_{-7.106}^{+16.95}$ & $4.398_{-2.882}^{+5.981}$ & $4.130_{-2.636}^{+7.605}$ & $4.889_{-3.224}^{+10.03}$ \\
 &  mean & $183.6$ & $20.49$ & $15.26$ & $6.458$ & $6.670$ & $9.555$ \\
 &  weighted & $386.6$ & $55.50$ & $30.70$ & $14.57$ & $15.09$ & $55.40$ \\
   \noalign{\smallskip}
\multirow{3}{*}{$\Sigma_{\rm SFR}$ thr $90$\%} &  median & $0.1043_{-0.0828}^{+0.3555}$ & $0.0056_{-0.0044}^{+0.0160}$ & $0.0055_{-0.0037}^{+0.0102}$ & $0.0026_{-0.0016}^{+0.0033}$ & $0.0025_{-0.0015}^{+0.0052}$ & $0.0028_{-0.0018}^{+0.0059}$ \\
 &  mean & $0.2510$ & $0.0147$ & $0.0098$ & $0.0041$ & $0.0047$ & $0.0070$ \\
 &  weighted & $0.5872$ & $0.0570$ & $0.0209$ & $0.0093$ & $0.0107$ & $0.0674$ \\
   \noalign{\smallskip}
\multirow{3}{*}{$\tau_{\rm dep}$ thr $90$\%} &  median & $1.177_{-0.697}^{+0.362}$ & $2.146_{-1.046}^{+1.720}$ & $1.798_{-0.725}^{+1.064}$ & $1.677_{-0.878}^{+1.377}$ & $1.564_{-0.747}^{+1.245}$ & $1.670_{-0.827}^{+1.320}$ \\
 &  mean & $1.138$ & $2.480$ & $2.017$ & $2.485$ & $2.245$ & $2.321$ \\
 &  weighted & $0.948$ & $2.310$ & $1.933$ & $2.567$ & $2.110$ & $2.135$ \\
  \hline
 \hline
 \multirow{3}{*}{$\Sigma_{\rm mol}$ -- $\alpha_{\rm CO}^{\rm MW}$} &  median & $163.6_{-126.7}^{+366.4}$ & $14.43_{-11.23}^{+39.21}$ & $9.715_{-6.524}^{+17.60}$ & $4.312_{-2.975}^{+7.189}$ & $3.768_{-2.519}^{+8.619}$ & $4.924_{-3.409}^{+11.85}$ \\
 &  mean & $251.6$ & $28.31$ & $15.96$ & $6.835$ & $6.883$ & $11.64$ \\
 &  weighted & $552.3$ & $84.21$ & $35.10$ & $16.85$ & $18.48$ & $110.5$ \\
   \noalign{\smallskip}
\multirow{3}{*}{$\Sigma_{\rm SFR}$ -- $\alpha_{\rm CO}^{\rm MW}$} &  median & $0.0739_{-0.0592}^{+0.3406}$ & $0.0057_{-0.0045}^{+0.0171}$ & $0.0055_{-0.0037}^{+0.0100}$ & $0.0026_{-0.0016}^{+0.0034}$ & $0.0025_{-0.0015}^{+0.0052}$ & $0.0029_{-0.0018}^{+0.0061}$ \\
 &  mean & $0.2079$ & $0.0156$ & $0.0094$ & $0.0042$ & $0.0046$ & $0.0074$ \\
 &  weighted & $0.5030$ & $0.0628$ & $0.0209$ & $0.0101$ & $0.0115$ & $0.0931$ \\
   \noalign{\smallskip}
\multirow{3}{*}{$t_{\rm dep}$ -- $\alpha_{\rm CO}^{\rm MW}$} &  median & $1.875_{-1.050}^{+0.660}$ & $2.548_{-1.228}^{+2.174}$ & $1.861_{-0.804}^{+0.868}$ & $1.675_{-0.911}^{+1.224}$ & $1.537_{-0.837}^{+1.174}$ & $1.709_{-0.912}^{+1.250}$ \\
 &  mean & $1.844$ & $3.084$ & $2.008$ & $2.209$ & $2.114$ & $2.227$ \\
 &  weighted & $1.539$ & $3.107$ & $2.028$ & $2.310$ & $2.138$ & $2.303$ \\
   \hline
   \noalign{\smallskip}
 \multirow{3}{*}{$\Sigma_{\rm mol}$ -- $\alpha_{\rm CO}^{\rm N12}$} &  median & $87.23_{-55.65}^{+86.73}$ & $25.30_{-16.70}^{+36.89}$ & $19.01_{-11.06}^{+19.62}$ & $12.42_{-7.212}^{+13.72}$ & $10.77_{-5.941}^{+13.33}$ & $13.49_{-7.904}^{+17.73}$ \\
 &  mean & $109.1$ & $34.67$ & $23.54$ & $16.01$ & $14.05$ & $19.43$ \\
 &  weighted & $174.9$ & $64.52$ & $36.78$ & $26.84$ & $22.33$ & $42.97$ \\
   \noalign{\smallskip}
\multirow{3}{*}{$\Sigma_{\rm SFR}$ -- $\alpha_{\rm CO}^{\rm N12}$} &  median & $0.1043_{-0.0828}^{+0.3103}$ & $0.0065_{-0.0049}^{+0.0180}$ & $0.0064_{-0.0041}^{+0.0102}$ & $0.0039_{-0.0020}^{+0.0045}$ & $0.0036_{-0.0019}^{+0.0063}$ & $0.0042_{-0.0024}^{+0.0077}$ \\
 &  mean & $0.2110$ & $0.0183$ & $0.0105$ & $0.0059$ & $0.0061$ & $0.0101$ \\
 &  weighted & $0.4052$ & $0.0481$ & $0.0184$ & $0.0105$ & $0.0104$ & $0.0405$ \\
   \noalign{\smallskip}
\multirow{3}{*}{$t_{\rm dep}$ -- $\alpha_{\rm CO}^{\rm N12}$} &  median & $0.945_{-0.550}^{+1.319}$ & $3.737_{-1.884}^{+2.993}$ & $2.886_{-1.192}^{+1.442}$ & $3.070_{-1.318}^{+1.793}$ & $2.736_{-1.178}^{+1.837}$ & $2.925_{-1.303}^{+1.905}$ \\
 &  mean & $1.184$ & $4.322$ & $3.107$ & $3.338$ & $3.063$ & $3.274$ \\
 &  weighted & $0.860$ & $3.631$ & $2.858$ & $3.439$ & $3.049$ & $3.105$ \\
  \hline
 \multirow{3}{*}{$\Sigma_{\rm mol}$ -- $\alpha_{\rm CO}^{\rm B13}$} &  median & $31.97_{-18.73}^{+17.05}$ & $18.12_{-8.696}^{+13.26}$ & $15.98_{-7.995}^{+11.72}$ & $15.25_{-7.713}^{+10.98}$ & $12.27_{-5.883}^{+6.921}$ & $14.33_{-7.070}^{+10.44}$ \\
 &  mean & $33.38$ & $21.66$ & $17.94$ & $16.89$ & $12.87$ & $16.27$ \\
 &  weighted & $45.14$ & $33.82$ & $23.96$ & $22.61$ & $16.06$ & $23.13$ \\
   \noalign{\smallskip}
\multirow{3}{*}{$\Sigma_{\rm SFR}$ -- $\alpha_{\rm CO}^{\rm B13}$} &  median & $0.0694_{-0.0547}^{+0.1495}$ & $0.0070_{-0.0053}^{+0.0156}$ & $0.0059_{-0.0037}^{+0.0098}$ & $0.0039_{-0.0023}^{+0.0052}$ & $0.0036_{-0.0019}^{+0.0060}$ & $0.0042_{-0.0024}^{+0.0076}$ \\
 &  mean & $0.1351$ & $0.0184$ & $0.0098$ & $0.0064$ & $0.0058$ & $0.0096$ \\
 &  weighted & $0.2310$ & $0.0419$ & $0.0150$ & $0.0090$ & $0.0078$ & $0.0199$ \\
   \noalign{\smallskip}
\multirow{3}{*}{$t_{\rm dep}$ -- $\alpha_{\rm CO}^{\rm B13}$} &  median & $0.454_{-0.275}^{+0.515}$ & $2.664_{-1.539}^{+2.899}$ & $2.560_{-1.150}^{+2.045}$ & $3.595_{-1.721}^{+2.580}$ & $2.958_{-1.516}^{+2.477}$ & $2.991_{-1.531}^{+2.546}$ \\
 &  mean & $0.549$ & $3.651$ & $3.033$ & $4.250$ & $3.535$ & $3.639$ \\
 &  weighted & $0.435$ & $3.135$ & $2.718$ & $4.207$ & $3.534$ & $3.447$ \\
   \hline
     \hline
\end{tabular}
\label{table:stats_appendix}
\end{center}
\end{table*}

In some cases, the ALMA maps cover a slightly larger or smaller fraction of the galaxy discs (e.g.\ relative to a characteristic radius such as $R_{25}$), and this different coverage could potentially bias the results that we presented. To confirm this, Table~\ref{table:stats_appendix} lists equivalent measurements to Table~\ref{table:stats} but restricted to galaxies with uniform azimuthal ALMA coverage out to $R_{25}$. The relative contribution of discs without spiral masks to molecular gas mass and SFR increases by a few percent (at the expense of the other environments dropping slightly, particularly the centres), but other than this the change is not dramatic.

We also consider another sanity check in Table~\ref{table:stats_appendix}. In Sect.~\ref{Sec:surfdens} we explained how kpc-size hexagonal apertures are assigned to a given environment if at least $80$\% of the high-resolution CO emission and SFR in the aperture falls within the footprint of an environment. Table~\ref{table:stats_appendix} shows an alternative to Table~\ref{table:stats_surfdens} varying this threshold from $70$\% to $90$\%. Increasing the threshold means that we are tossing out a larger number of `mixed' apertures, in order to focus on a smaller subset of apertures that are more uniquely associated with a given environment. The changes on the median $\Sigma_{\rm mol}$, $\Sigma_{\rm SFR}$, and $\tau_{\rm dep}$ are typically a few percent, and only very rarely exceed $10$\% when we perturb this threshold within the range $70$\% to $90$\%. Some of these changes are expected: for example, increasing the threshold means that we only retain the innermost apertures for some centre masks, and, consequently, the surface densities tend to increase slightly, but never dramatically. Therefore, we can conclude that the choice of threshold does not strongly impact our results.

\onecolumn
\begin{center}
\begin{longtable}{lccccccc}
\caption{Galaxy sample studied in this paper. \label{table:sample}}\\ 
\hline\hline 
& log($M_\star$/$M_\odot$) & log(SFR/[$M_\odot$\,yr$^{-1}$]) & $V_{\rm min}$/[km\,s$^{-1}$] & $V_{\rm max}$/[km\,s$^{-1}$] & centre & bar & spiral  \\ 
\hline
\endfirsthead
\caption{Continued.} \\
\hline\hline
& log($M_\star$/$M_\odot$) & log(SFR/[$M_\odot$\,yr$^{-1}$]) & $V_{\rm min}$/[km\,s$^{-1}$] & $V_{\rm max}$/[km\,s$^{-1}$] & centre & bar & spiral  \\ 
\hline
\endhead
\hline
\endfoot
IC1954     &    9.7 &  -0.44 &    900 &   1175 &   1 &   1 &   1 \\
IC5273     &    9.7 &  -0.27 &   1150 &   1400 &   1 &   1 &   0 \\
IC5332     &    9.7 &  -0.39 &    625 &    750 &   1 &   0 &   0 \\
NGC0628    &   10.3 &   0.24 &    600 &    700 &   1 &   0 &   1 \\
NGC0685    &   10.1 &  -0.38 &   1225 &   1450 &   0 &   1 &   0 \\
NGC1087    &    9.9 &   0.12 &   1375 &   1625 &   0 &   1 &   0 \\
NGC1097    &   10.8 &   0.68 &   1000 &   1550 &   1 &   1 &   1 \\
NGC1300    &   10.6 &   0.07 &   1400 &   1700 &   1 &   1 &   1 \\
NGC1317    &   10.6 &  -0.32 &   1825 &   2025 &   1 &   1 &   0 \\
NGC1365    &   11.0 &   1.23 &   1400 &   1850 &   1 &   1 &   1 \\
NGC1385    &   10.0 &   0.32 &   1350 &   1600 &   0 &   0 &   1 \\
NGC1433    &   10.9 &   0.05 &    925 &   1175 &   1 &   1 &   0 \\
NGC1511    &    9.9 &   0.36 &   1150 &   1475 &   1 &   0 &   0 \\
NGC1512    &   10.7 &   0.11 &    725 &   1025 &   1 &   1 &   1 \\
NGC1546    &   10.4 &  -0.08 &   1050 &   1450 &   1 &   0 &   0 \\
NGC1559    &   10.4 &   0.58 &   1100 &   1450 &   0 &   1 &   0 \\
NGC1566    &   10.8 &   0.66 &   1325 &   1650 &   1 &   1 &   1 \\
NGC1637    &    9.9 &  -0.19 &    600 &    825 &   1 &   1 &   1 \\
NGC1672    &   10.7 &   0.88 &   1200 &   1450 &   1 &   1 &   1 \\
NGC1792    &   10.6 &   0.57 &   1050 &   1375 &   1 &   0 &   0 \\
NGC1809    &    9.8 &   0.76 &   1125 &   1400 &   1 &   0 &   0 \\
NGC2090    &   10.0 &  -0.39 &    700 &   1100 &   1 &   0 &   1 \\
NGC2283    &    9.9 &  -0.28 &    725 &    925 &   1 &   1 &   1 \\
NGC2566    &   10.7 &   0.94 &   1475 &   1775 &   1 &   1 &   1 \\
NGC2775    &   11.1 &  -0.06 &   1100 &   1575 &   1 &   0 &   0 \\
NGC2835    &   10.0 &   0.09 &    750 &    975 &   1 &   1 &   1 \\
NGC2903    &   10.6 &   0.49 &    300 &    775 &   1 &   1 &   0 \\
NGC2997    &   10.7 &   0.64 &    925 &   1225 &   1 &   0 &   1 \\
NGC3059    &   10.4 &   0.38 &   1150 &   1325 &   0 &   1 &   0 \\
NGC3137    &    9.9 &  -0.31 &    950 &   1225 &   1 &   0 &   0 \\
NGC3239    &    9.2 &  -0.41 &    648 &    848 &   0 &   0 &   0 \\
NGC3351    &   10.4 &   0.12 &    625 &    950 &   1 &   1 &   0 \\
NGC3507    &   10.4 &  -0.00 &    875 &   1075 &   1 &   1 &   1 \\
NGC3511    &   10.0 &  -0.09 &    925 &   1250 &   1 &   1 &   0 \\
NGC3521    &   11.0 &   0.57 &    550 &   1050 &   1 &   0 &   0 \\
NGC3596    &    9.7 &  -0.52 &   1100 &   1275 &   1 &   0 &   0 \\
NGC3621    &   10.1 &  -0.00 &    575 &    875 &   1 &   0 &   0 \\
NGC3626    &   10.5 &  -0.67 &   1200 &   1675 &   1 &   1 &   0 \\
NGC3627    &   10.8 &   0.58 &    500 &    950 &   1 &   1 &   1 \\
NGC4207    &    9.7 &  -0.72 &    480 &    720 &   0 &   0 &   0 \\
NGC4254    &   10.4 &   0.49 &   2250 &   2525 &   1 &   0 &   1 \\
NGC4293    &   10.5 &  -0.29 &    775 &   1075 &   1 &   1 &   0 \\
NGC4298    &   10.0 &  -0.34 &   1000 &   1275 &   1 &   0 &   0 \\
NGC4303    &   10.5 &   0.73 &   1450 &   1675 &   1 &   1 &   1 \\
NGC4321    &   10.7 &   0.55 &   1425 &   1725 &   1 &   1 &   1 \\
NGC4424    &    9.9 &  -0.52 &    350 &    550 &   0 &   0 &   0 \\
NGC4457    &   10.4 &  -0.51 &    786 &    986 &   1 &   1 &   0 \\
NGC4496A   &    9.5 &  -0.21 &   1625 &   1825 &   0 &   1 &   0 \\
NGC4535    &   10.5 &   0.33 &   1800 &   2100 &   1 &   1 &   1 \\
NGC4536    &   10.4 &   0.54 &   1600 &   2000 &   1 &   1 &   1 \\
NGC4540    &    9.8 &  -0.78 &   1200 &   1375 &   1 &   1 &   0 \\
NGC4548    &   10.7 &  -0.28 &    350 &    650 &   1 &   1 &   1 \\
NGC4569    &   10.8 &   0.12 &   -425 &      0 &   1 &   1 &   0 \\
NGC4571    &   10.1 &  -0.54 &    225 &    450 &   1 &   0 &   0 \\
NGC4579    &   11.1 &   0.34 &   1300 &   1725 &   1 &   1 &   1 \\
NGC4654    &   10.6 &   0.58 &    850 &   1225 &   0 &   1 &   0 \\
NGC4689    &   10.2 &  -0.39 &   1500 &   1800 &   1 &   0 &   0 \\
NGC4694    &    9.9 &  -0.81 &   1060 &   1300 &   0 &   0 &   0 \\
NGC4731    &    9.5 &  -0.22 &   1375 &   1600 &   0 &   1 &   1 \\
NGC4781    &    9.6 &  -0.32 &   1100 &   1400 &   0 &   1 &   0 \\
NGC4826    &   10.2 &  -0.69 &    200 &    625 &   1 &   0 &   0 \\
NGC4941    &   10.2 &  -0.35 &    900 &   1300 &   1 &   1 &   0 \\
NGC4951    &    9.8 &  -0.45 &   1025 &   1350 &   1 &   0 &   0 \\
NGC5042    &    9.9 &  -0.22 &   1250 &   1525 &   1 &   0 &   0 \\
NGC5068    &    9.4 &  -0.56 &    625 &    725 &   1 &   1 &   0 \\
NGC5128    &   11.0 &   0.09 &    200 &    850 &   1 &   0 &   0 \\
NGC5134    &   10.4 &  -0.34 &   1625 &   1860 &   1 &   1 &   0 \\
NGC5248    &   10.4 &   0.36 &   1000 &   1325 &   1 &   1 &   1 \\
NGC5530    &   10.1 &  -0.48 &   1025 &   1350 &   1 &   0 &   0 \\
NGC5643    &   10.3 &   0.41 &   1050 &   1325 &   1 &   1 &   1 \\
NGC6300    &   10.5 &   0.28 &    900 &   1300 &   1 &   1 &   0 \\
NGC6744    &   10.7 &   0.38 &    625 &   1025 &   1 &   1 &   1 \\
NGC7456    &    9.6 &  -0.43 &   1050 &   1310 &   1 &   0 &   0 \\
NGC7496    &   10.0 &   0.35 &   1525 &   1750 &   1 &   1 &   0 \\
\end{longtable}
\tablefoot{\mod{Stellar mass and star formation rates of the galaxies studied in this paper \citep{2021arXiv210407739L}. $V_{\rm min}$ and $V_{\rm max}$ indicate the velocity range (LSRK) over which the ALMA cubes were integrated in order to obtain the intensity maps used in this paper. The last three columns indicate the morphological components included in the environmental masks. All galaxies with a spiral mask have a corresponding interarm mask, whereas the rest have a generic disc component.}}
\end{center}

\twocolumn

\begin{figure*}[t]
\begin{center}
\includegraphics[trim=0 0 0 0, clip,width=0.85\textwidth]{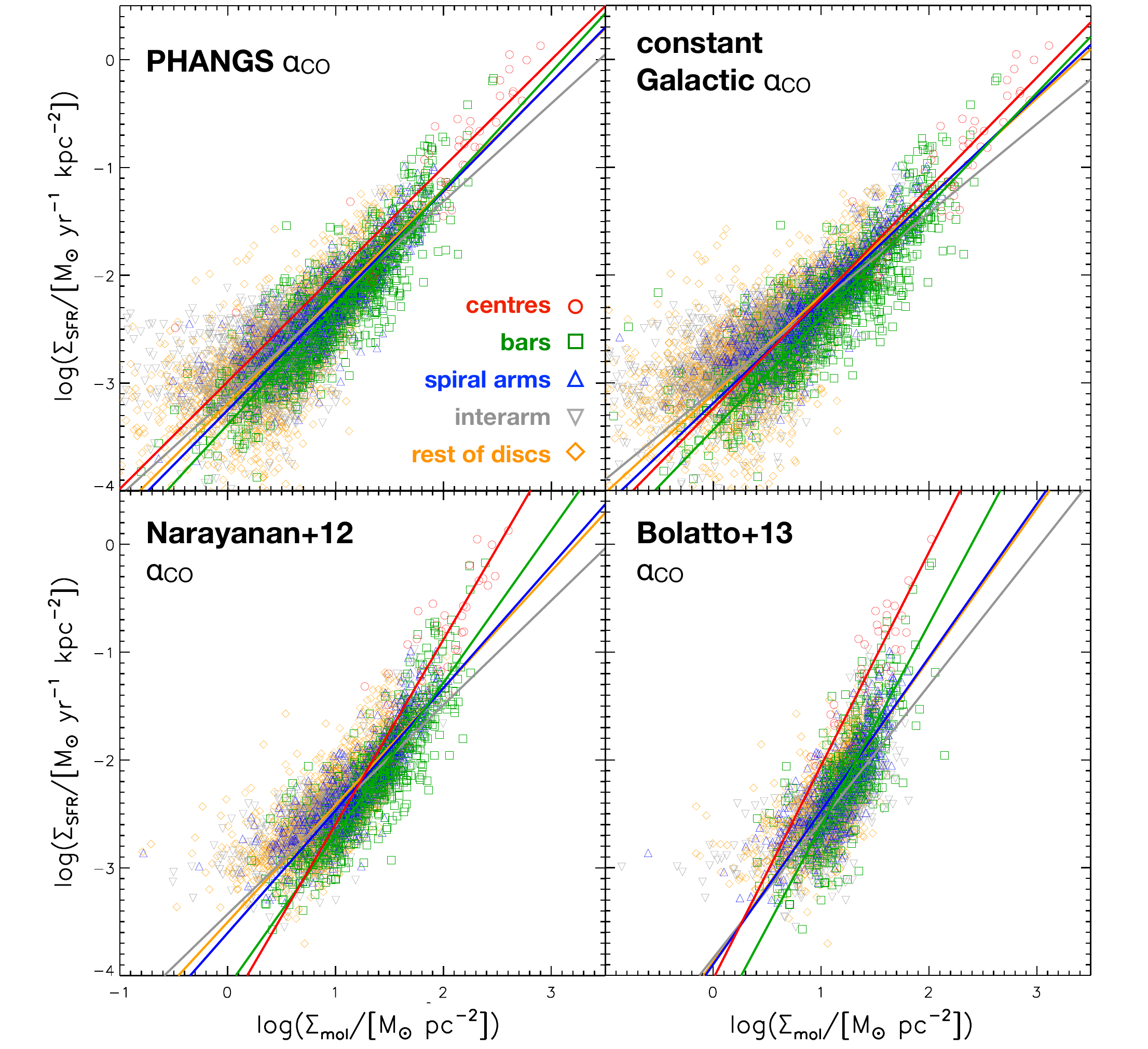}
\end{center}
\caption{Molecular Kennicutt--Schmidt relation for kpc-scale measurements in PHANGS--ALMA using alternative prescriptions for the $\alpha_{\rm CO}$ conversion factor, as indicated in the top-left of each panel. The straight lines represent the best bisector fit to the data for the different environments, as indicated by the different colours. The numerical results of the bisector fits are listed in Table~\ref{table:KSfits}.
}
\label{fig:KSplot_alphaCO}
\end{figure*}

Finally, \mod{the bottom part of Table~\ref{table:stats_appendix} and Fig.~\ref{fig:KSplot_alphaCO} show the properties and molecular Kennicutt--Schmidt relation on kpc scales for alternative prescriptions of the $\alpha_{\rm CO}$ conversion factor. The numerical results of the Kennicutt--Schmidt} fits are listed in Table~\ref{table:KSfits}. 
For each $\alpha_{\rm CO}$ prescription, we consider only simultaneous detections in both surface densities (i.e.\ $\Sigma_{\rm mol}>0$, $\Sigma_{\rm SFR}>0$). A constant Galactic conversion factor ($4.35$\,$M_\odot$\,pc$^{-2}$\,(K\,km\,s$^{-1})^{-1}$; \citealt{2013ARA&A..51..207B}), which is a relatively widespread choice, results in slopes that are very similar to the ones that we find for our preferred PHANGS $\alpha_{\rm CO}$, but they tend to be slightly smaller (i.e.\ minimally less linear). The largest discrepancy arises if we follow the prescription from \citetalias{2013ARA&A..51..207B}, with a conversion factor that depends on the local cloud-scale molecular gas surface density, metallicity, and the average disc surface densities at kpc scales, including the gas and stellar components. This results in considerably steeper power-laws, but also reduces the dynamical range in molecular gas surface densities and the fits have large uncertainties. \mod{We note that the plots and fits for the \citetalias{2013ARA&A..51..207B} prescription only show a subset of 51 galaxies (out of 74), as {\sc H\,i} data are not yet available for all PHANGS galaxies. Similarly, for the \citetalias{2012MNRAS.421.3127N} recipe, three galaxies were discarded due to lack of CO data at 150\,pc resolution (given the newest distance estimates).}

\section{Atlas}
\label{sec:Appendix2}

\begin{figure*}[t]
\begin{center}
\includegraphics[trim=0 20 0 10, clip,width=0.95\textwidth]{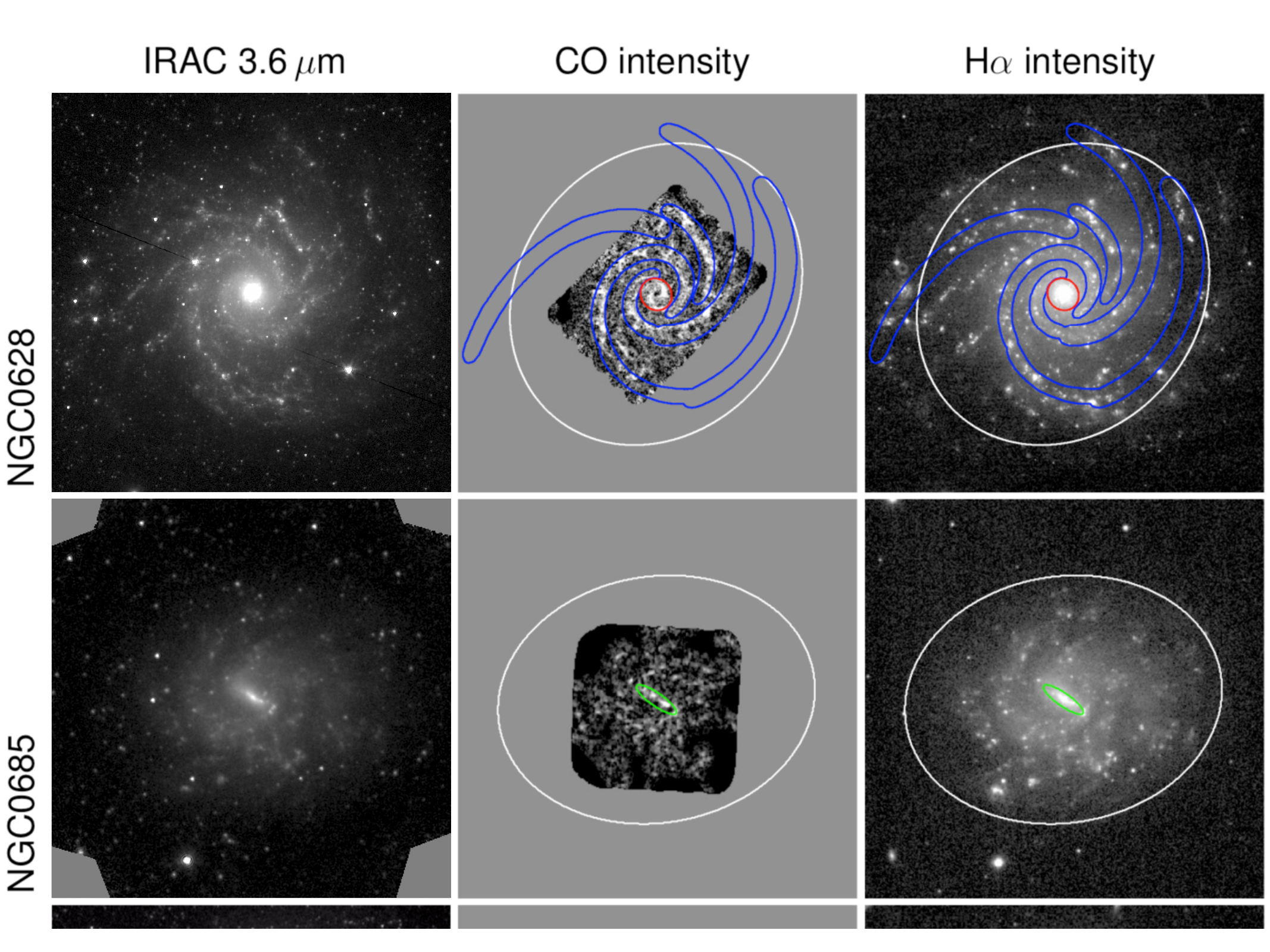}
\end{center}
\caption{First two rows of the online-only figure showing the atlas of galaxies with environmental masks. {\it Left}: {\it Spitzer} $3.6$\,$\mu$m images, in a square root stretch ([$0,\,2.5$] MJy\,sr$^{-1}$). The environmental masks are defined based on this NIR emission. {\it Middle}: \mbox{CO(2--1)} emission (strict moment map combining ALMA 12\,m and 7\,m arrays, as well as Total Power); shown in square root stretch ([$0,\,10$] K\,km\,s$^{-1}$). {\it Right}: H$\alpha$ emission, shown in square root stretch. The middle and right panels are overlaid with contours for the environments, with the same colour-coding as in the paper (red = centre; green = bar; blue = spiral arms; \mod{white} = galaxy disc).
}
\label{fig:atlas}
\end{figure*}

Figure\,\ref{fig:atlas} is available as supplementary online material, and contains an atlas of the main PHANGS--ALMA sample of galaxies showing the footprint of environments. This includes the {\it Spitzer} $3.6$\,$\mu$m images, tracing \mod{stellar mass}, and which the environmental masks are based on. Next to it, we show \mbox{CO(2--1)} emission tracing molecular gas, as well as  H$\alpha$ emission, tracing star formation. The contours highlight the various environments, with the same colour-coding as used in the rest of the paper.

\bibliography{mq}{}
\bibliographystyle{aa}{}

\end{document}